\documentclass[3p,11pt]{elsarticle}

\usepackage{amssymb}
\usepackage{amsmath}
\usepackage[T1]{fontenc}
\usepackage{subcaption}
\usepackage[section]{placeins}
\usepackage{booktabs}
\usepackage{url}
\usepackage[hidelinks]{hyperref}

\journal{Applied Energy}

\begin{document}

\begin{frontmatter}

\title{Compound effects of traffic and climate on electric vehicle HVAC energy consumption: a spatiotemporal framework with city-level attribution}

\fntext[conf]{The short version of the paper was presented at ICAE2025, Bangkok, Thailand, Dec.~8--12, 2025. This paper is a substantial extension of the short version of the conference paper.}

\author[kcl]{Liang Zhang}
\ead{liang.zhang@kcl.ac.uk}

\author[kcl]{Wei He\corref{cor1}}
\cortext[cor1]{Corresponding author}
\ead{wei.4.he@kcl.ac.uk}

\affiliation[kcl]{organization={Department of Engineering, King's College London},
            city={London},
            country={United Kingdom}}

\begin{abstract}
Electric vehicle (EV) energy consumption in real-world driving departs significantly from rated values due to the joint influence of ambient temperature, traffic congestion, and route characteristics. Existing studies typically evaluate these factors in isolation or under static conditions, leaving the compound effect of co-varying climate and traffic on HVAC energy unquantified and the relative contribution of each factor to route-level variability unknown. This study develops a spatiotemporal simulation framework that, for the first time, couples traffic-aware driving speed, location- and time-specific ambient temperature, and physics-based energy submodels (cabin HVAC, traction, and battery thermal management) at the segment level, and introduces a regression-based decomposition to attribute HVAC energy variability to its temperature and trip-duration components on a per-route basis.

The framework is applied through a factorial experimental design across seven UK cities and eight radial routes, systematically varying temporal factors (departure time, intra-day traffic) and spatial factors (regional climate, road-network type). Results show that while total energy consumption varies by 14\% across cities, the HVAC component varies by up to 89\%, identifying cabin thermal management as the primary differentiator between routes under winter conditions. The per-route decomposition reveals that trip duration, determined by traffic and road type, is frequently the dominant driver of HVAC energy variability: in London, 83\% of the above-average HVAC energy is attributable to congestion-extended trip time rather than temperature. The decomposition yields a closed-form prediction model that estimates HVAC energy from only three inputs (ambient temperature, average speed, and trip distance), offering physical interpretability, negligible computational cost, and straightforward transferability to other vehicles and regions through re-fitting of three coefficients. These findings suggest that EV range variability is substantially shaped by traffic conditions and road-network characteristics, providing actionable guidance for route planning, infrastructure design, and energy equity assessment.
\end{abstract}


\end{frontmatter}

\section{Introduction}

Electric vehicles (EVs) are central to decarbonising the transport sector. Governments worldwide have set ambitious adoption targets, including the UK banning new petrol and diesel car sales from 2030 \cite{CMA} and China mandating EVs as the dominant vehicle type before 2030 under its New Energy Vehicle Industry Development Plan \cite{Ministry}, while market projections suggest EV penetration in private vehicle ownership could exceed 90\% by 2050 \cite{calvillo2020analysing}. Yet real-world EV energy consumption can deviate by 20--40\% from rated values \cite{yuksel2015effects,AAA2019EVRange}, driven by external factors that conventional range estimates do not capture.

Among these factors, cabin heating, ventilation, and air conditioning (HVAC) consistently emerges as one of the largest contributors to the gap between rated and actual range \cite{sanguinetti2017many}. Cold climates increase energy consumption by up to 24\% \cite{neubauer2014thru}, and extreme temperatures (both cold and hot) can reduce range by 29--36\% \cite{yuksel2015effects,farrington2000impact}. The magnitude of the penalty is sensitive to operating conditions: cabin pre-conditioning can recover up to 10\% of range \cite{kambly2014estimating}, implying that HVAC demand is not a fixed overhead but a dynamic, route-dependent load. Because HVAC energy depends jointly on climate and traffic conditions, EV users in cold, congested areas may face systematically higher energy costs per kilometre, an emerging dimension of transport energy inequality \cite{bauer2021might,DOE2024_cold_BEV} that current range estimates do not address.

Despite the recognised impact of these factors, existing studies predominantly evaluate them in isolation or under static conditions \cite{nguyen2019energy}. Temperature effects are typically assessed at fixed driving speeds; traffic impacts are analysed without accounting for thermal loads; and HVAC energy is often modelled as a constant auxiliary draw rather than a dynamic, route-dependent quantity. Few studies have examined how these factors interact when they vary simultaneously across time and space \cite{yuksel2015effects,hendricks2001vehicle}. This separation has practical consequences. Range estimation tools that treat HVAC as a static overhead cannot explain why the same vehicle on the same route consumes substantially more energy during morning rush hour than at midday, or why two cities with comparable climates can differ markedly in HVAC energy intensity. Without a coupled treatment, the gap between rated and actual range remains largest precisely where EV adoption is most needed: in dense, congested urban environments.

Two critical gaps follow. First, no existing framework propagates traffic-aware driving speed and location-specific ambient temperature jointly through a coupled thermal-kinematic simulation at the segment level, such that the speed profile simultaneously determines traction energy and HVAC integration time. Without this coupling, models cannot capture the compound penalty that arises when congestion extends the duration over which thermal loads must be sustained. Second, while individual factor sensitivities have been reported, no study has quantitatively decomposed the relative contributions of temperature and traffic (trip duration) to HVAC energy variability across routes, cities, and seasons. Without this decomposition, range variability is treated as a monolithic climate phenomenon, and policy interventions risk being misdirected: a city might invest in heat pump subsidies when congestion relief would yield greater energy savings, or allocate charging infrastructure based on temperature zones rather than on actual per-kilometre energy demand.

This study addresses these gaps by developing a spatiotemporal simulation framework and applying it through a factorial experimental design to quantify the compound and separable effects of temporal and spatial factors on EV energy consumption. The specific contributions are:

\begin{enumerate}
    \item \textbf{Coupled spatiotemporal framework.} A segment-level simulation that integrates traffic-aware speed profiles, location- and time-specific ambient temperature, and physics-based energy submodels (cabin HVAC, traction, battery thermal management) to compute route-level energy consumption under realistic, dynamic conditions. By propagating thermal and kinematic states jointly along a route, the framework captures compound effects that static or decoupled approaches miss.
    \item \textbf{Factorial experimental design.} Three complementary experiments (diurnal variation, cross-city urban loops, and radial trips from a single origin) that systematically span the space--time domain and enable controlled comparison of factor effects. In particular, the radial experiment exploits near-constant temperature across directions from a single inland origin, isolating the road-network contribution to HVAC energy for the first time.
    \item \textbf{Decomposition of HVAC energy.} A regression-based attribution method that separates the contributions of ambient temperature and trip duration (determined by traffic and road type) to HVAC energy variability on a per-route basis.
    \item \textbf{Practical prediction model.} Derivation of a parsimonious closed-form model for HVAC energy as a function of ambient temperature, average speed, and trip distance. Unlike data-driven alternatives that require training data and offer no causal insight, this model provides physically interpretable coefficients suitable for direct integration into route planning and range estimation tools.
\end{enumerate}

\section{Methodology}

\subsection{Spatiotemporal Framework Overview}

Real-world EV energy consumption is strongly influenced by two classes of external factors that vary simultaneously: \emph{temporal operational factors} such as traffic-determined driving speed and time-dependent ambient temperature, and \emph{spatial factors} such as regional climate, road network type (urban versus motorway), and route direction. This work couples them through a segment-level simulation framework in which the vehicle's thermal and kinematic states evolve jointly along a route as functions of both location and time.

The framework links three components in a causal chain:
\begin{center}
\emph{traffic-aware speed profile} $\rightarrow$ \emph{spatiotemporal temperature field} $\rightarrow$ \emph{coupled energy-consumption models}.
\end{center}
At each simulation time step, the vehicle position is advanced according to the traffic-aware speed; the ambient temperature is queried from a spatiotemporal database using the current time and location; and the cabin thermal state, battery thermal state, and energy consumption are updated accordingly. Total energy is decomposed into traction, cabin HVAC, and battery thermal management (BTMS) components, enabling attribution of route-level differences to specific physical mechanisms.

To isolate and recombine the effects of temporal and spatial factors, the framework is applied through a factorial experimental design comprising three complementary experiments:

\begin{table}[htbp]
\centering
\caption{Factorial experimental design: systematic variation of temporal and spatial factors.}
\label{tab:factorial}
\small
\begin{tabular*}{\textwidth}{@{\extracolsep{\fill}}llll@{}}
\hline
Experiment & Spatial & Temporal & Factor isolated \\
\hline
Diurnal (London) & Fixed (one city) & Varied (departure hour) & Temporal (traffic + temperature) \\
City loops (7 cities) & Varied (across UK) & Fixed (08:00) & Compound (climate + traffic) \\
Radial (Manchester) & Varied (8 directions) & Fixed (08:00) & Spatial (road network) \\
\hline
\end{tabular*}
\end{table}

The diurnal experiment isolates temporal effects by fixing the spatial context. The city loop experiment reveals the compound effect of co-varying climate and urban traffic. The radial experiment holds temperature approximately constant ($<$1\textdegree C range) while varying route direction and road type, thereby isolating the spatial road-network effect. Results from the latter two experiments are pooled into a unified regression to decompose HVAC energy into temperature and trip-duration contributions.

\subsection{Vehicle Energy Simulation Platform}

The simulation builds on the open-source vehicle energy platform FASTSim (Future Automotive Systems Technology Simulator) \cite{brooker2015fastsim}, developed by the U.S. National Renewable Energy Laboratory (NREL). FASTSim performs second-by-second vehicle energy flow simulation and supports various powertrain architectures, but its default framework focuses on traction energy and propulsion subsystems.

In this work, FASTSim computes traction energy consumption using detailed vehicular parameters and traffic-derived driving cycles. On top of this baseline, a dynamic cabin thermal balance model is integrated externally to represent HVAC loads. This model accounts for heat exchange between the cabin and the ambient environment, the thermal mass of the cabin, and the controller response that drives the cabin temperature toward the comfort setpoint. As a result, HVAC electrical power naturally exhibits an initially high demand when the cabin-to-setpoint temperature difference is large, followed by a gradual decrease as the cabin approaches equilibrium. A battery thermal management model is similarly coupled to capture temperature-dependent battery conditioning loads.

\subsection{Cabin Thermal Dynamics and HVAC Energy Model}

\subsubsection{Cabin 1R1C Lumped Thermal Model}

To balance physical interpretability and computational efficiency, the cabin thermal process is approximated by a first-order lumped model (1R1C), in which the entire cabin air volume is treated as a single thermal node governed by one state variable, the cabin air temperature $T_{cab}$. Heat exchange with the ambient temperature is represented by an effective thermal resistance (or, equivalently, an overall heat transfer coefficient). The cabin energy balance is written as

\begin{equation}
C_{cab}\frac{dT_{cab}}{dt} = UA_{cab}\left(T_{amb}-T_{cab}\right) + Q_{hvac},
\label{eq:cabin_1r1c}
\end{equation}

where $T_{amb}$ is the ambient temperature, $C_{cab}$ (J/K) is the equivalent cabin thermal capacity, $UA_{cab}$ (W/K) is the equivalent overall heat transfer coefficient (aggregating conduction and convection pathways), and $Q_{hvac}$ (W) is the net thermal power delivered by the HVAC system (positive for heating; negative for cooling). The corresponding effective thermal resistance is

\begin{equation}
R_{th,cab}=\frac{1}{UA_{cab}}.
\label{eq:rth_ua}
\end{equation}

A key advantage of the 1R1C structure is its strong interpretability. It naturally separates transient warm-up/cool-down phases from near-steady maintenance phases, which enables segment-level attribution of HVAC energy (e.g., cold-start conditioning versus steady holding).

The proposed system-level modelling paradigm aligns with FASTSim’s LumpedCabin/HVAC framework, which formulates the cabin as a \textit{lumped thermal node}, and applies an HVAC control law. Within this formulation, the temperature setpoint, deadband, controller gain, and HVAC power limit are specified to evaluate thermal heat flow and the corresponding energy consumption.

\subsubsection{HVAC Thermal-Power Command: Feedforward + Proportional Correction with Deadband and Saturation}

Given a temperature setpoint $T_{set}$, the commanded HVAC thermal power is constructed as the sum of a feedforward term and a proportional correction term, followed by actuator saturation:

\begin{equation}
Q_{cmd}=\mathrm{clip}\!\left(Q_{ff}+Q_{p},\,-Q_{\max},\,Q_{\max}\right),
\label{eq:qcmd}
\end{equation}
where $Q_{\max}$ denotes the maximum deliverable HVAC thermal power and $\mathrm{clip}(\cdot)$ enforces physical output limits. The command comprises two components. The feedforward term compensates the quasi-steady heat-transfer load:
\begin{equation}
Q_{ff}=UA_{cab}\left(T_{set}-T_{amb}\right),
\label{eq:qff}
\end{equation}
providing the baseline thermal power required to maintain the setpoint when $T_{cab}\approx T_{set}$ and thereby reducing steady-state bias. A proportional correction addresses transient tracking and modelling mismatch:
\begin{equation}
Q_{p}=K_{p}\left(T_{set}-T_{cab}\right),
\label{eq:qp_basic}
\end{equation}
where $K_p$ is the proportional gain. To prevent control chattering near the setpoint, a deadband $\Delta T_{db}$ disables the proportional term for small temperature errors:
\begin{equation}
Q_p =
\begin{cases}
0, & |T_{set}-T_{cab}| < \Delta T_{db},\\
K_p(T_{set}-T_{cab}), & \text{otherwise}.
\end{cases}
\label{eq:qp_deadband}
\end{equation}
Finally, actuator saturation constrains the commanded thermal power to the feasible operating envelope,
\begin{equation}
Q_{cmd} \in [-Q_{\max},\,Q_{\max}],
\end{equation}
ensuring physical realism under extreme ambient conditions. In saturation regimes, the available HVAC capacity may be insufficient to fully track $T_{set}$, leading to persistent temperature offsets that the model captures for realistic energy and comfort assessments. Together, this formulation makes the causal chain \emph{temperature difference} $\rightarrow$ \emph{thermal load} $\rightarrow$ \emph{commanded thermal power} $\rightarrow$ \emph{electrical energy} explicit, enabling spatiotemporal HVAC energy decomposition at segment level.

\subsubsection{COP-Based Mapping from Thermal Power to Electrical Power}

The electrical power consumption of the HVAC system is determined by the dynamic thermal load and the coefficient of performance (COP). The ideal COP follows from the Carnot cycle: $\mathrm{COP}_{\mathrm{ideal}} = T_{\mathrm{cabin}} / (T_{\mathrm{ambient}} - T_{\mathrm{cabin}})$ in cooling mode and $\mathrm{COP}_{\mathrm{ideal}} = T_{\mathrm{ambient}} / (T_{\mathrm{cabin}} - T_{\mathrm{ambient}})$ in heating mode. Due to system inefficiencies (compressor losses, heat exchanger effectiveness), the actual COP is corrected as $\mathrm{COP}_{\mathrm{actual}} = \mathrm{COP}_{\mathrm{ideal}} \times \eta_{\mathrm{cop}}$, where $\eta_{\mathrm{cop}} = 0.18$ is consistent with typical values in the FASTSim documentation \cite{peri2017cool}. The instantaneous electrical power required by the HVAC system is then
\[
P_{\mathrm{hvac}}(t) = \frac{Q_{\mathrm{cmd}}(t)}{\mathrm{COP}_{\mathrm{actual}}(t)},
\]
where $Q_{\mathrm{cmd}}(t)$ is provided by the dynamic cabin heat balance model rather than a static assignment, capturing the transient decrease of HVAC demand as the cabin temperature converges toward the setpoint.

\subsection{Traction Energy Model}

\subsubsection{Road-Load Forces and Wheel Power}
A quasi-steady longitudinal road-load model is used. Rolling resistance and aerodynamic drag are
\begin{equation}
F_{roll}=mgC_{rr},\qquad
F_{aero}=\frac{1}{2}\rho C_dA\,v^2,
\label{eq:forces}
\end{equation}
where $m$ is vehicle mass, $g$ gravitational acceleration, $C_{rr}$ rolling resistance coefficient, $\rho$ air density, $C_dA$ effective drag area, and $v$ vehicle speed. The wheel-end power is
\begin{equation}
P_{wheel}=(F_{roll}+F_{aero})v.
\label{eq:pwheel}
\end{equation}

\subsubsection{Battery-Side Power and Segment Traction Energy}
Battery-side traction power is approximated using a constant drivetrain efficiency $\eta_{drv}$ plus a baseline auxiliary load:
\begin{equation}
P_{batt}=\frac{P_{wheel}}{\eta_{drv}}+P_{aux,base}.
\label{eq:pbatt}
\end{equation}
For a segment (or time step) with actual used duration $\Delta t_{used}$ (in seconds), the traction energy is
\begin{equation}
E_{trac}=\frac{P_{batt}\Delta t_{used}}{3.6\times 10^{6}}\quad(\mathrm{kWh}),
\label{eq:etrac}
\end{equation}
using $1~\mathrm{kWh}=3.6\times 10^6~\mathrm{J}$.

\paragraph{Role in the coupled framework.}
The traction model provides a speed-driven baseline (notably $F_{aero}\propto v^2$) and supports cross-vehicle transfer through parameterisation by $(m,C_dA,C_{rr},\eta_{drv},P_{aux,base})$.

\subsection{Battery Thermal Management System (BTMS)}

Following the established energy-balance approach \cite{bernardi1985_energy_balance,togun2025_btms_review}, a first-order lumped battery thermal model is adopted:
\begin{equation}
C_{batt}\frac{dT_{batt}}{dt}=Q_{gen}+Q_{btms}-UA_{batt}(T_{batt}-T_{amb}),
\label{eq:batt_thermal}
\end{equation}
where $T_{batt}$ is battery-pack temperature, $C_{batt}$ the equivalent thermal capacitance, $UA_{batt}$ the equivalent heat-transfer coefficient to ambient, $Q_{gen}$ internal heat generation, and $Q_{btms}$ the BTMS delivered thermal power (positive for heating; negative for cooling) \cite{ghaeminezhad2023_control_oriented_review,lumped_lp_preferred_realtime}.

A rule-based deadband controller tracks a target temperature $T_{target}$: heating is activated when $T_{batt}$ falls below the band (limited by $Q_{heat,\max}$), and cooling when above (limited by $Q_{cool,\max}$) \cite{nam2025_ann_mpc_rulebased_baseline}. Electrical power is computed as
\begin{equation}
P_{btms}=\frac{|Q_{btms}|}{COP_{heat/cool}}+P_{aux,btms},
\label{eq:pbtms}
\end{equation}
where $COP_{heat/cool}$ denotes the effective COP and $P_{aux,btms}$ denotes auxiliary loads \cite{hao2024_itms_vcc_review}. Separating BTMS energy from cabin HVAC enables explicit ``thermal share'' attribution at the route level.

\subsection{Spatiotemporal Temperature Database}

\subsubsection{Design Requirement}
Under the conditions of equal distance but different routes and/or different departure times, HVAC and BTMS energy depend on the ambient temperature encountered along the route. Hence, a data source is required to provide ambient temperature as a function of time and location:
\begin{equation}
T_{amb}=T_{amb}(t,\mathbf{x}).
\end{equation}

\subsubsection{Establishment of a Temperature Database}

To support high-resolution thermal simulations, we developed a local temperature database implemented in SQLite that stores hourly ambient temperature data for the United Kingdom. Temperature time series are retrieved from the Renewables.ninja API \cite{pfenninger2016pv,staffell2016wind}, which provides MERRA-2 reanalysis data on a regular latitude-longitude grid with a spatial resolution of $0.625^{\circ} \times 0.5^{\circ}$. To ensure query consistency and eliminate redundant downloads, each queried city or route waypoint is snapped to the corresponding MERRA-2 grid cell prior to data requests, and the resulting hourly records are cached for subsequent reuse. In addition to the primary temperature table, we construct a representative day's table to identify season-specific representative dates (e.g., typical, colder-than-average, and warmer-than-average) from the full-year dataset. During simulation, the model queries the local database using the EV's timestamp and geolocation to obtain the corresponding ambient temperature input.

\subsection{Traffic-Aware Routing and Segment-Wise Coupled Integration}

\subsubsection{Traffic-Aware Speed Profile from Routing Services}
Route speed and traffic conditions are obtained using a routing API (e.g., Google Routes) under a specified departure Time. A field mask is used to request only necessary fields to reduce response size and latency. Conceptually, the routing output defines a ``speed database'': the speed experienced when the vehicle, departing at a given time, reaches each route segment or location.

\subsubsection{Coupled Temperature Queries and State Propagation}
During simulation, the vehicle position is advanced in fixed-time steps according to a traffic-aware speed profile. At each step, the ambient temperature is queried from the local temperature database using the concurrent time $(t,\mathbf{x})$, and the cabin and battery thermal states $(T_{cab},T_{batt})$ are updated. Energy is accumulated via time integration.

\subsubsection{Unified Energy Decomposition and Route-Level Metrics}
For time step (or segment) $k$ with average speed $v_k$, distance $d_k$, duration of time used $\Delta t_{used,k}$, and ambient temperature $T_{amb,k}$, the total energy is decomposed as
\begin{equation}
E_{total,k}=E_{trac,k}+E_{hvac,k}+E_{btms,k},
\label{eq:etotal}
\end{equation}
and the route total is
\begin{equation}
E_{route}=\sum_k E_{total,k}.
\label{eq:eroute}
\end{equation}
Average energy intensity is reported as
\begin{equation}
\mathrm{eff}=\frac{E_{route}}{\sum_k d_k}\times 100\quad(\mathrm{kWh}/100\,\mathrm{km}).
\label{eq:eff}
\end{equation}

To interpret route differences, component shares are also reported. For example,
\begin{align}
\mathrm{HVAC\_share} &= \frac{\sum_k E_{hvac,k}}{E_{route}},\\
\mathrm{BTMS\_share} &= \frac{\sum_k E_{btms,k}}{E_{route}},\\
\mathrm{Thermal\_share} &= \frac{\sum_k (E_{hvac,k}+E_{btms,k})}{E_{route}}.
\end{align}

\subsubsection{Cross-Thermal-Zone Integration}
When a trip traverses multiple thermal zones, the framework avoids representing the entire route with a single mean ambient temperature. Instead, the routing service returns segment-level kinematics $(v_k, d_k, \Delta t_k)$, and the temperature database provides the corresponding $T_{\mathrm{amb},k}$ for each segment via its location and timestamp. Cabin and battery thermal states $(T_{\mathrm{cab}}, T_{\mathrm{batt}})$ are updated sequentially across segments, capturing transient effects such as cold-start warm-up and steady-state maintenance.

\subsection{Vehicle Parameterisation via External Configuration}
Vehicle and thermal-control parameters are supplied through an external YAML configuration file, and loaded at the simulation entry point. This enables straightforward replacement of vehicle models within the same framework, supporting systematic comparisons across vehicles, routes, and departure times.

In this study, the Tesla Model Y is selected as the representative vehicle for simulation, and the collected vehicle configuration parameters are presented in Table \ref{tab:configuration}. The data are taken from Tesla's published specifications for the 2025 Model Y RWD

\begin{table}[htbp]
    \centering
    \caption{Vehicle configuration parameters}
    \label{tab:configuration}
    \begin{tabular}{llcc}
        \toprule
        Category & Parameter & Value & Unit \\
        \midrule
        {Vehicle dynamics}
            & Vehicle mass & 1900 & kg \\
            & Maximum motor power & 220 & kW \\
            & Transmission efficiency & 0.98 & -- \\
            & Base auxiliary power & 250 & W \\
        \midrule
        {Battery system}
            & Battery energy capacity & 62.5 & kWh \\
            & Maximum thermal-limited power & 15 & kW \\
            & Internal resistance & 1.0 & $\Omega$ \\
        \midrule
        {HVAC / thermal model}
            & Cabin heat capacitance & $1.8\times10^{5}$ & J/K \\
            & Conductance to ambient & 55 & W/K \\
            & Conductance to cabin & 1 & W/K \\
            & HVAC maximum thermal power & 15000 & W \\
            & Auxiliary HVAC power (cabin) & 5000 & W \\
            & Auxiliary HVAC power (battery) & 5000 & W \\
            & COP correction factor $\eta_{\mathrm{cop}}$ & 0.18 & -- \\
        \bottomrule
    \end{tabular}
\end{table}

\section{Model Validation}

\subsection{Validation Purpose}

The energy model comprises two principal subsystems: a steady-state traction model and a 1R1C lumped-cabin HVAC model with a Carnot-fraction COP formulation. The purpose of this validation is to verify that: (1)~the traction subsystem correctly captures the speed--energy relationship; (2)~the HVAC subsystem correctly captures the temperature--heating power relationship; and (3)~the combined prediction produces physically consistent total energy consumption across temperatures. The validation draws on three independent data sources covering three vehicles, temperatures from $-10$ to $38\,^\circ$C, and speeds from 46.5 to 130~km/h.

For comparative route analysis, systematic biases are acceptable because they cancel when computing energy differences between routes at the same temperature. What matters is that the model preserves the correct ranking and relative magnitude of differences caused by speed, temperature, and route geometry.

\subsection{Validation Vehicles and Data Sources}

\begin{itemize}
    \item[\textbf{(a)}] \textbf{Tesla Model~Y LR AWD.} Constant-speed road tests by Bj{\o}rn Nyland~\cite{Nyland2021,Nyland2022} at 90 and 120~km/h in summer (20\,$^\circ$C) and winter (5\,$^\circ$C), and by ArenaEV~\cite{ArenaEV2023} at 90 and 130~km/h (38\,$^\circ$C). The AWD shares the same platform, body, battery, and heat pump as the target RWD. These six data points directly verify traction and HVAC parameters for the Model~Y across a wide speed and temperature range.

    \item[\textbf{(b)}] \textbf{Tesla Model~3 LR AWD 2022 (heat pump).} Subsystem-level cross validation. Tested on a chassis dynamometer at TEEL, Ottawa~\cite{Humphries2026}. Component-level energy decomposition at $-10$, $-7$, 0, and 25\,$^\circ$C using Hioki PW6001 power analysers enables independent validation of traction, COP, HVAC power, and cold-start energy. SAE~J1634 SMCT+ procedure with UDDS (34~km/h), HWFET (78~km/h), and CSC65 (105~km/h) cycles. Shares the same platform and heat pump architecture as the Model~Y.

    \item[\textbf{(c)}] \textbf{BMW i3 60\,Ah (PTC heater).} Independent cross-platform verification. Real-world 10\,Hz CAN-bus data from Steinstraeter et al.~\cite{Steinstraeter2020}. The HVAC heating system is based on a PTC heater instead of a heat pump.
\end{itemize}

\begin{table}[htbp]
    \centering
    \caption{Specifications of the validation vehicles and the target simulation vehicle.}
    \label{tab:val-specs}
    \small
    \begin{tabular*}{\textwidth}{@{\extracolsep{\fill}}lcccc@{}}
    \hline
    Parameter & Model~Y LR AWD & Model~3 LR AWD 2022 & BMW i3 60\,Ah\\
    \hline
    Heating system  & Heat pump      & Heat pump (R1234yf) & PTC resistive \\
    Mass (kg)       & 1979           & 1838                & 1245          \\
    $C_dA$ (m$^2$)  & $\sim$0.62     & 0.511               & 0.690         \\
    $C_{rr}$        & 0.008          & 0.007               & 0.010         \\
    Battery (kWh)   & 75             & 82.1                & 18.2          \\
    Data type       & Road (const.\ speed) & Chassis dyno  & Road (10\,Hz CAN) \\
    Source          & \cite{Nyland2021,Nyland2022,ArenaEV2023} & \cite{Humphries2026} & \cite{Steinstraeter2020} \\
    \hline
    \end{tabular*}
\end{table}

\subsection{Validation Results}
\label{sec:val-results}

\subsubsection{Traction Energy}
\label{sec:val-traction}

Traction energy is predicted within $\pm 10$\% across all three vehicles, spanning a mass range of 1245--1979~kg, $C_dA$ range of 0.511--0.690~m$^2$, and speeds from 46.5 to 130~km/h (Table~\ref{tab:val-traction}). Ten data points from four independent sources yield a mean absolute percentage error (MAPE) of 6.3\%.

\begin{table}[htbp]
    \centering
    \caption{Traction energy validation across three vehicles.}
    \label{tab:val-traction}
    \small
    \begin{tabular*}{\textwidth}{@{\extracolsep{\fill}}llccccc@{}}
    \hline
    Vehicle & Source & Speed (km/h) & Measured (kWh/100km) & Model & Error & $T_\mathrm{amb}$ ($^\circ$C) \\
    \hline
    Model~Y  & Bj{\o}rn Nyland  & 90   & 14.2 & 14.9 & $+5.1$\%  & 20 \\
    Model~Y  & Bj{\o}rn Nyland  & 120  & 20.6 & 21.8 & $+5.9$\%  & 20 \\
    Model~Y  & Bj{\o}rn Nyland  & 90   & 15.5 & 16.1 & $+4.0$\%  & 5  \\
    Model~Y  & Bj{\o}rn Nyland  & 120  & 21.4 & 23.3 & $+8.7$\%  & 5  \\
    Model~Y  & ArenaEV          & 90   & 14.1 & 14.4 & $+2.2$\%  & 38 \\
    Model~Y  & ArenaEV          & 130  & 22.2 & 23.5 & $+5.7$\%  & 38 \\
    Model~3  & TEEL          & 105  & 14.1 & 14.9 & $+5.8$\%  & 25 \\
    Model~3  & TEEL          & 78   & 11.6 & 10.6 & $-8.8$\%  & 25 \\
    BMW i3   & Steinstraeter    & 72   & 10.4 & 11.4 & $+9.6$\%  & 23 \\
    \hline
    \end{tabular*}
\end{table}

\begin{figure}[!htbp]
  \centering
  \includegraphics[width=0.55\textwidth]{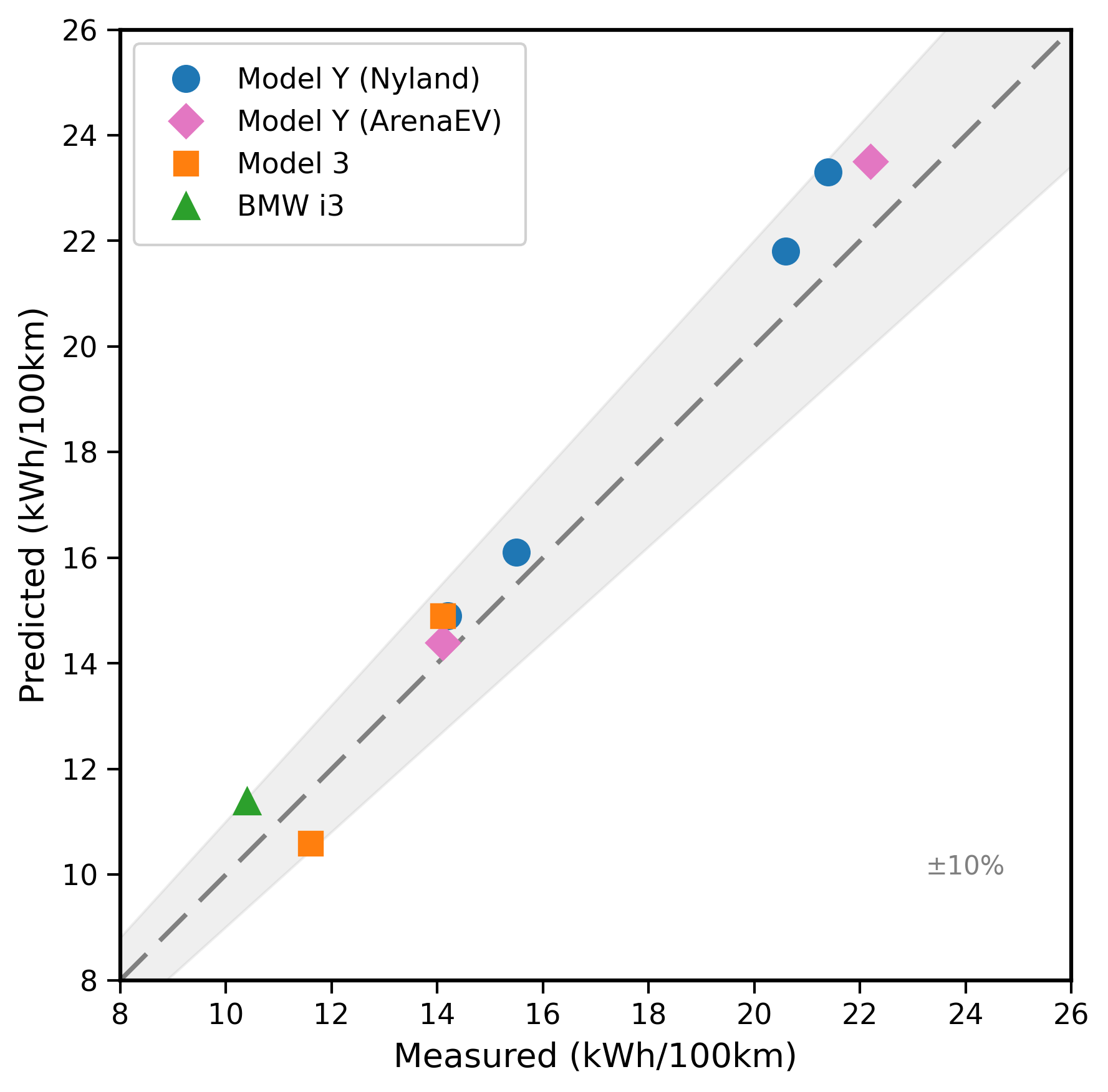}
  \caption{Traction energy predicted versus measured comparison}
  \label{fig:val-traction-scatter}
\end{figure}

Predicted and measured values are compared in Fig.~\ref{fig:val-traction-scatter}; all ten points fall within $\pm 10$\%.

 The Model~Y winter 120~km/h point ($+8.7$\%) shows the largest positive error, partly because the test data include moderate acceleration, braking, and stop-start events that dissipate kinetic energy; this accounts for the slightly larger errors at these conditions compared with the constant-speed tests.

Critically, the model correctly reproduces the magnitudes of both the speed effect and the temperature effect on the target Model~Y:

\begin{description}
    \item[Speed effect:] Bj{\o}rn Nyland measured $+6.4$~kWh/100km ($+45$\%) from 90 to 120~km/h at 20\,$^\circ$C; the model predicts $+6.9$ ($+46$\%). The near-identical slope confirms that the aerodynamic and rolling resistance parameters correctly capture the Model~Y's speed--energy relationship.
    \item[Temperature effect:] At 90~km/h, the measured increase from 20\,$^\circ$C to 5\,$^\circ$C is $+1.3$~kWh/100km ($+9.2$\%); the model predicts $+1.2$ ($+8.1$\%). This close agreement in temperature-driven increments directly validates the model's ability to quantify seasonal energy differences on the target vehicle.
\end{description}

\subsubsection{HVAC Steady-State Power}
\label{sec:val-hvac-ss}

Under steady-state conditions (cabin stabilised at 22\,$^\circ$C), predicted HVAC electrical power matches measured values with a MAPE of 9.5\% (Table~\ref{tab:val-hvac}).

\begin{table}[htbp]
    \centering
    \caption{HVAC steady-state power validation.}
    \label{tab:val-hvac}
    \small
    \begin{tabular*}{\textwidth}{@{\extracolsep{\fill}}lcccccr@{}}
    \hline
    Vehicle & $T_\mathrm{amb}$ ($^\circ$C) & Speed (km/h) & Measured (kW) & Model (kW) & COP & Error \\
    \hline
    Model~3 HP & $-10$ & 105 & 1.50 & 1.52 & 1.66 & $+1$\%  \\
    Model~Y RWD & $-7$  & 46.5 & 1.20 & 1.15 & 1.20 & $-4.2$\% \\
    i3 PTC     & 2.3   & 55  & 1.47 & 1.39 & 1.00 & $-5$\%  \\
    i3 PTC     & 2.6   & 54  & 1.28 & 1.36 & 1.00 & $+7$\%  \\
    \hline
    \end{tabular*}
\end{table}

\begin{figure}[!htbp]
  \centering
  \includegraphics[width=0.55\textwidth]{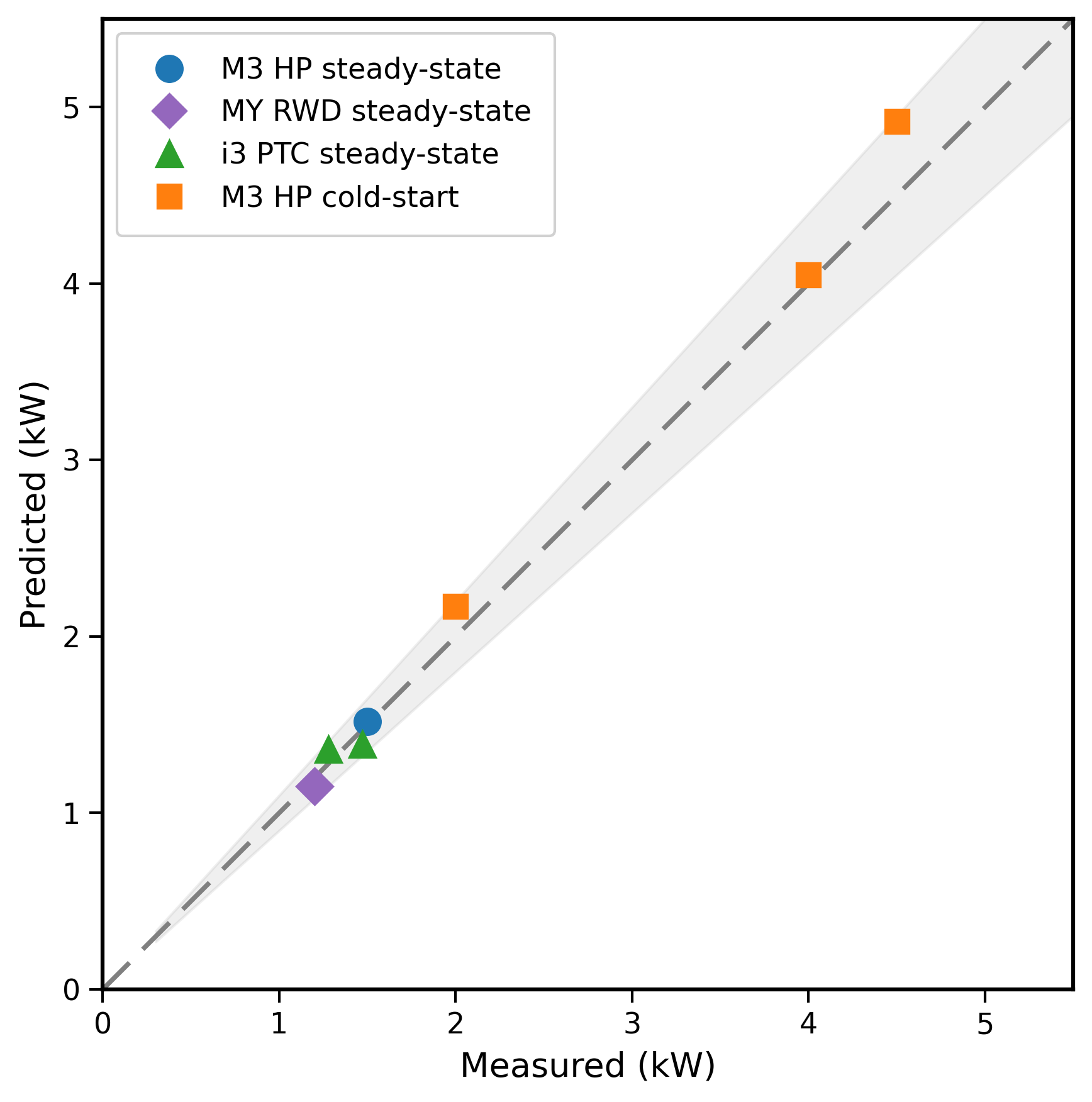}
  \caption{HVAC power predicted versus measured comparison}
  \label{fig:val-hvac-scatter}
\end{figure}

Across both steady-state heating (1--1.5~kW) and cold-start conditions (2--5~kW), all points fall within the $\pm 10$\% band (Fig.~\ref{fig:val-hvac-scatter}).

\subsubsection{Heat Pump Coefficient of Performance (COP)}

The COP is modelled as $\mathrm{COP} = \eta \cdot T_\mathrm{set} / (T_\mathrm{set} - T_\mathrm{amb})$ with $\eta = 0.18$, verified post hoc against TEEL measurements (Table~\ref{tab:val-cop}). The model COP falls within the measured range at all three temperatures.

\begin{table}[htbp]
    \centering
    \caption{Heat pump COP validation.}
    \label{tab:val-cop}
    \small
    \begin{tabular*}{\textwidth}{@{\extracolsep{\fill}}cccc@{}}
    \hline
    $T_\mathrm{amb}$ ($^\circ$C) & Model COP & Measured range & Within range \\
    \hline
    $-10$ & 1.66 & 0.80\,--\,1.93 & Yes \\
    $-7$  & 1.83 & 1.09\,--\,2.10 & Yes \\
    0     & 2.41 & 1.53\,--\,3.33 & Yes \\
    \hline
    \end{tabular*}
\end{table}

\subsubsection{HVAC Cold-Start Energy}

The cold-start term $E_\mathrm{cs} = C_\mathrm{eff} \times (T_\mathrm{set} - T_\mathrm{init}) / \mathrm{COP}$ accounts for warming the effective thermal mass ($C_\mathrm{eff} \approx 200$~kJ/K, including cabin, structure, and battery preheat). Table~\ref{tab:val-coldstart} compares with TEEL first-UDDS data (1370~s, cabin soaked at ambient). The MAPE is 9.3\%.

\begin{table}[htbp]
    \centering
    \caption{HVAC cold-start validation (Model~3 HP, $T_\mathrm{cab}$ initial $= T_\mathrm{amb}$).}
    \label{tab:val-coldstart}
    \small
    \begin{tabular*}{\textwidth}{@{\extracolsep{\fill}}ccccccc@{}}
    \hline
    $T_\mathrm{amb}$ ($^\circ$C) & Measured (kW) & $E_\mathrm{cs}$ (kWh) & $E_\mathrm{ss}$ (kWh) & Model (kW) & Error & $C_\mathrm{eff}$ (kJ/K) \\
    \hline
    $-10$ & 4.5 & 1.071 & 0.801 & 4.92 & $+9$\%  & 200 \\
    $-7$  & 4.0 & 0.879 & 0.663 & 4.05 & $+1$\%  & 200 \\
    0     & 2.0 & 0.506 & 0.389 & 2.17 & $+8$\% & 200 \\
    \hline
    \end{tabular*}
\end{table}

\subsection{Validation Summary and Implications for Route Analysis}

\begin{table}[htbp]
    \centering
    \caption{Validation summary. MAPE = mean absolute percentage error; Bias = mean signed error ($+$ = overestimation).}
    \label{tab:val-summary}
    \small
    \begin{tabular*}{\textwidth}{@{\extracolsep{\fill}}lcccc@{}}
    \hline
    Metric & $n$ & MAPE & Bias \\
    \hline
    Traction (3 vehicles)           & 10 & 6.3\%  & $+2.9$\%  \\
    Model~Y                          & 6  & 5.3\%  & $+5.3$\% \\
    HVAC steady-state power          & 4  & 9.5\%  & $+7.0$\% \\
    HVAC cold-start                  & 3  & 9.3\%  & $+9.3$\%  \\
    Heat pump COP                    & 3  & ---    & ---       \\
    Total energy                    & 7  & 8.0\%  & $-7.9$\%  \\
    \hline
    \end{tabular*}
\end{table}

The validation confirms the model is suitable for comparative route energy analysis:

\begin{enumerate}
    \item \textbf{Traction energy is predicted within $\pm 10$\% across three vehicles and ten conditions (MAPE 6.3\%)}, with the Model~Y achieving MAPE 5.3\% over six constant-speed points. The speed--energy slope (measured $+45$\% vs model $+46$\%) and temperature--energy increment (measured $+9.2$\% vs model $+8.1$\%) on the Model~Y are reproduced within 1~percentage point, directly validating the model's ability to capture the physical effects that drive route energy differences.

    \item \textbf{HVAC steady-state power (MAPE 9.5\%), cold-start energy (MAPE 9.3\%), and COP (3/3 within range)} confirm the temperature--HVAC relationship. The critical $-10\,^\circ$C steady-state point matches within $+1$\%.

    \item \textbf{Systematic biases in traction ($+4.2$\%) and HVAC ($+7.0$\%) are small and of the same sign}, cancelling when computing energy differences between routes at the same temperature.
\end{enumerate}

\paragraph{Limitations.}
(a)~The steady-state traction model does not capture kinetic energy losses from stop-start and acceleration/braking events, which contributes to the underprediction observed in drive cycles with frequent speed changes (HWFET, WLTP).
(b)~The cabin-only HVAC model does not separately represent battery preheating or defrost loads.

\section{Results}

The validated model is applied to real-world travel scenarios in the United Kingdom through the factorial experimental design introduced in Table~\ref{tab:factorial}. Results are presented in five stages: a baseline response surface establishes the two-input physics governing HVAC energy; the diurnal experiment isolates temporal effects within London; a cross-city comparison reveals the compound interaction of climate and traffic; radial routes from Manchester isolate the road-network effect; and a unified decomposition quantifies per-route attribution and yields a practical prediction model. Throughout, energy is disaggregated into traction, cabin HVAC, and battery thermal management (BTMS) components, with the thermal-system share defined as
\begin{equation}
\text{Thermal\_share}=\frac{E_{\text{hvac}}+E_{\text{btms}}}{E_{\text{total}}}.
\end{equation}

\subsection{HVAC energy response surface}

\begin{figure}[!htbp]
  \centering
  \includegraphics[width=\linewidth]{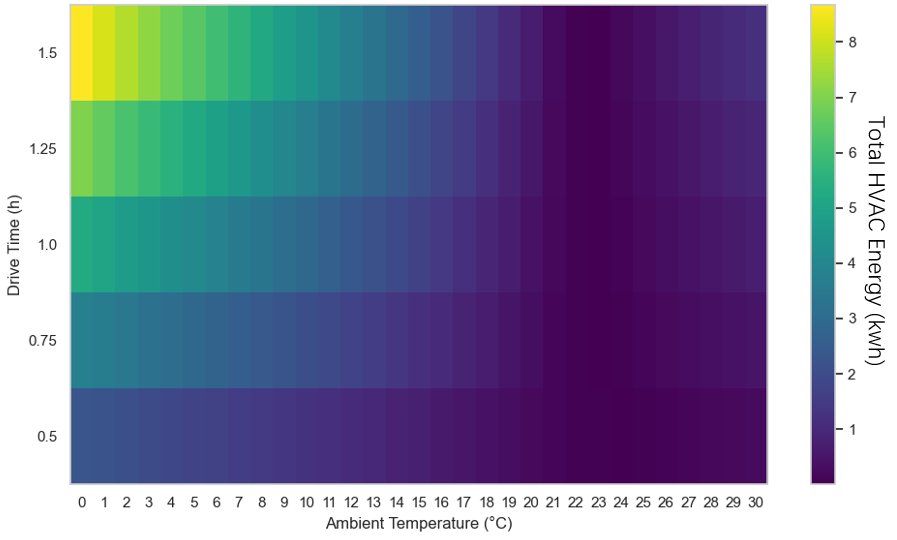}
  \caption{Distribution of HVAC energy consumption across different ambient temperatures.}
  \label{fig:2D}
\end{figure}

Among the factors affecting HVAC energy consumption, two dominate under the conditions examined here: ambient temperature, which sets the instantaneous thermal load, and trip duration, which determines how long that load persists. The response surface in Fig.~\ref{fig:2D} captures this two-input dependence: HVAC energy increases approximately linearly with driving time but exhibits a pronounced nonlinear dependence on ambient temperature, rising steeply as temperature deviates from the cabin setpoint. This surface serves as the organising frame for the results that follow: spatial effects (regional climate differences) shift routes horizontally across the temperature axis, while traffic effects (congestion-induced delays) shift routes vertically along the duration axis.

An immediate implication is that congestion carries a quantifiable thermal energy cost: at a given ambient temperature $T$, each unit of delay $\Delta t$ adds approximately $\bar{P}_{\text{hvac}}(T)\Delta t$ to the HVAC energy budget. Even modest congestion can therefore accumulate meaningful thermal energy over a full trip. The diurnal experiment in London tests this relationship directly.

\subsection{Diurnal variation in London: temporal effects on energy consumption}
\label{sec:diurnal}

\begin{figure}[!htbp]
  \centering
  \includegraphics[width=\linewidth]{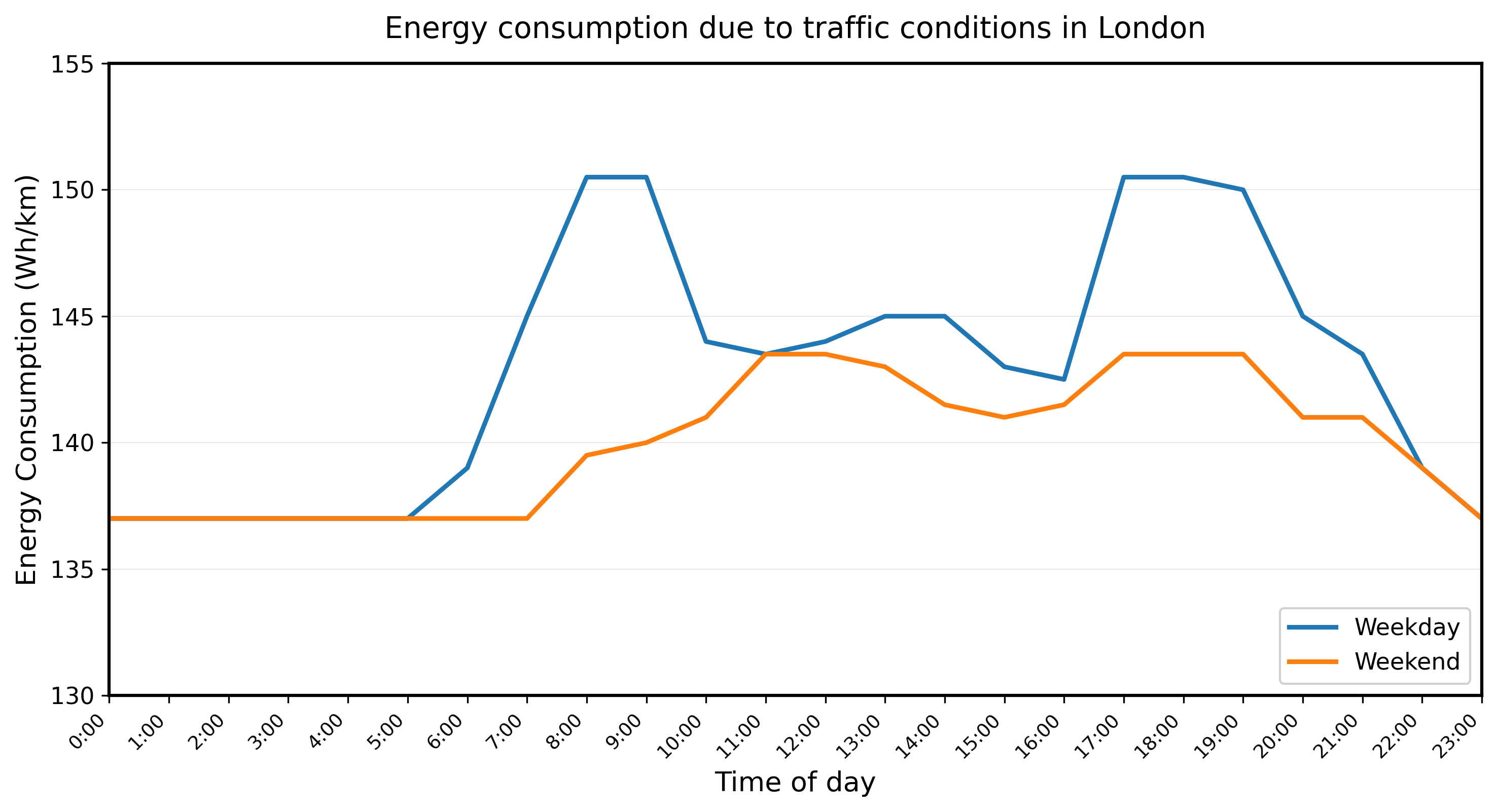}
  \caption{Time-dependent energy consumption under weekday and weekend conditions.}
  \label{fig:traffic}
\end{figure}

\begin{figure}[!htbp]
  \centering
  \includegraphics[width=\linewidth]{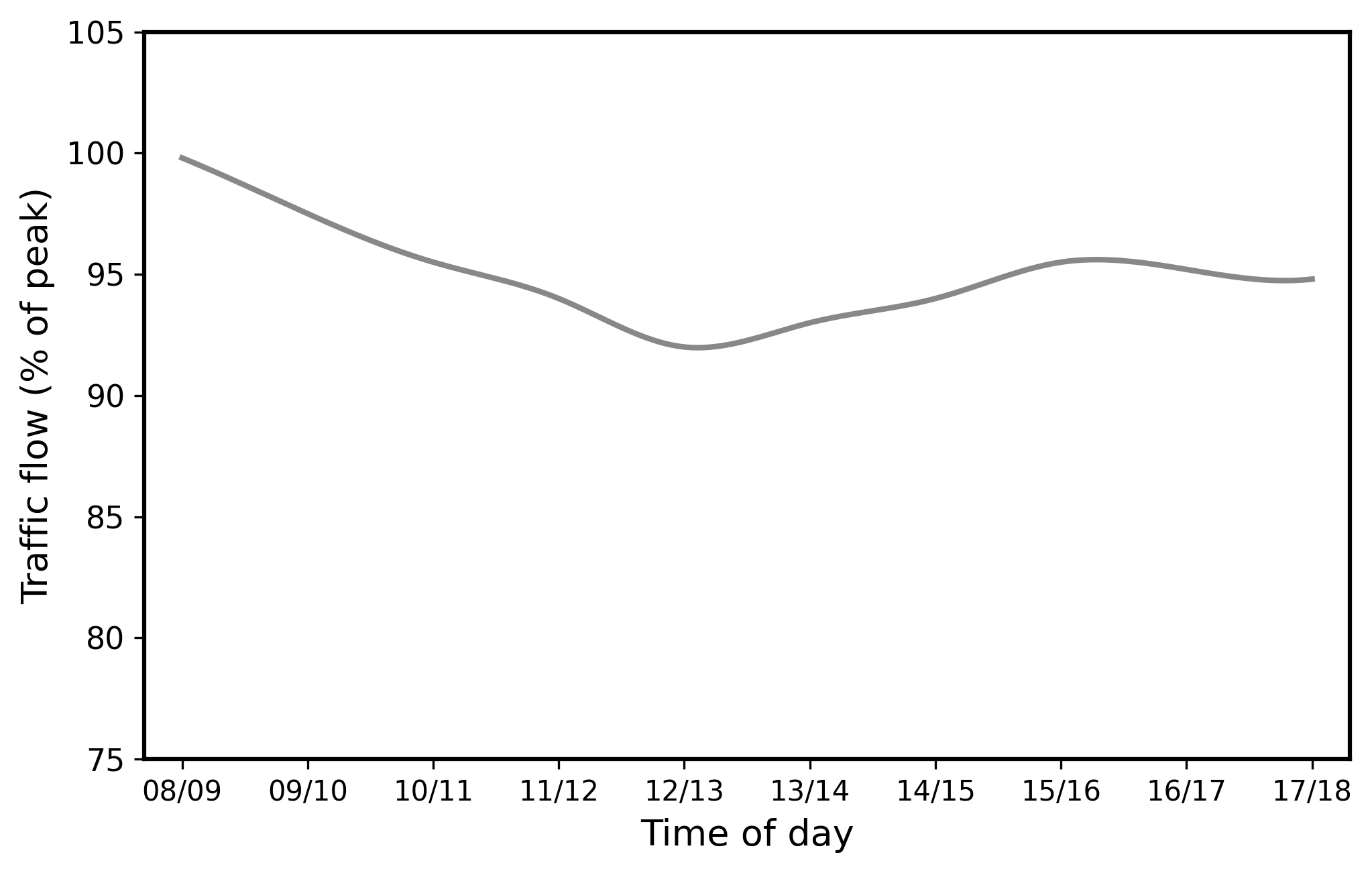}
  \caption{London 24-hour average weekday traffic flow.}
  \label{fig:Flow}
\end{figure}

Within a single city, departure time alone produces substantial energy variation. Simulated weekday departures from London exhibit clear morning and evening energy peaks that align closely with the Transport for London (TfL) 2024 intra-day congestion index \cite{TFL_2024_AnnualOverview} (Figs.~\ref{fig:traffic}--\ref{fig:Flow}), while weekend departures yield a flatter profile. Even under identical vehicle parameters and climate conditions, the choice of departure hour changes energy consumption measurably.

Decomposing the peak--off-peak differences into energy components,
\begin{equation}
\Delta E_{\text{total}}=\Delta E_{\text{trac}}+\Delta E_{\text{hvac}}+\Delta E_{\text{btms}},
\end{equation}
reveals the mechanism through which traffic amplifies thermal loads. HVAC \textit{power} profiles across departure times are broadly similar, because the ambient temperature within London varies only modestly over the course of a day. However, HVAC \textit{energy} differs substantially, because congestion extends the trip duration over which that power must be sustained; the time integral of a similar power curve over a longer window yields more energy. This distinction between instantaneous power and cumulative energy is central to interpreting all subsequent results.

Within London, temperature is approximately constant across departure times, so the observed temporal variation is driven predominantly by trip duration. To reveal the temperature effect, the analysis must move from a single city to a cross-city comparison in which regional climate co-varies with urban traffic conditions.

\subsection{Cross-city comparison: compound effects of climate and traffic}
\label{sec:cityloops}

\begin{figure}[!htbp]
  \centering
  \includegraphics[width=0.85\textwidth]{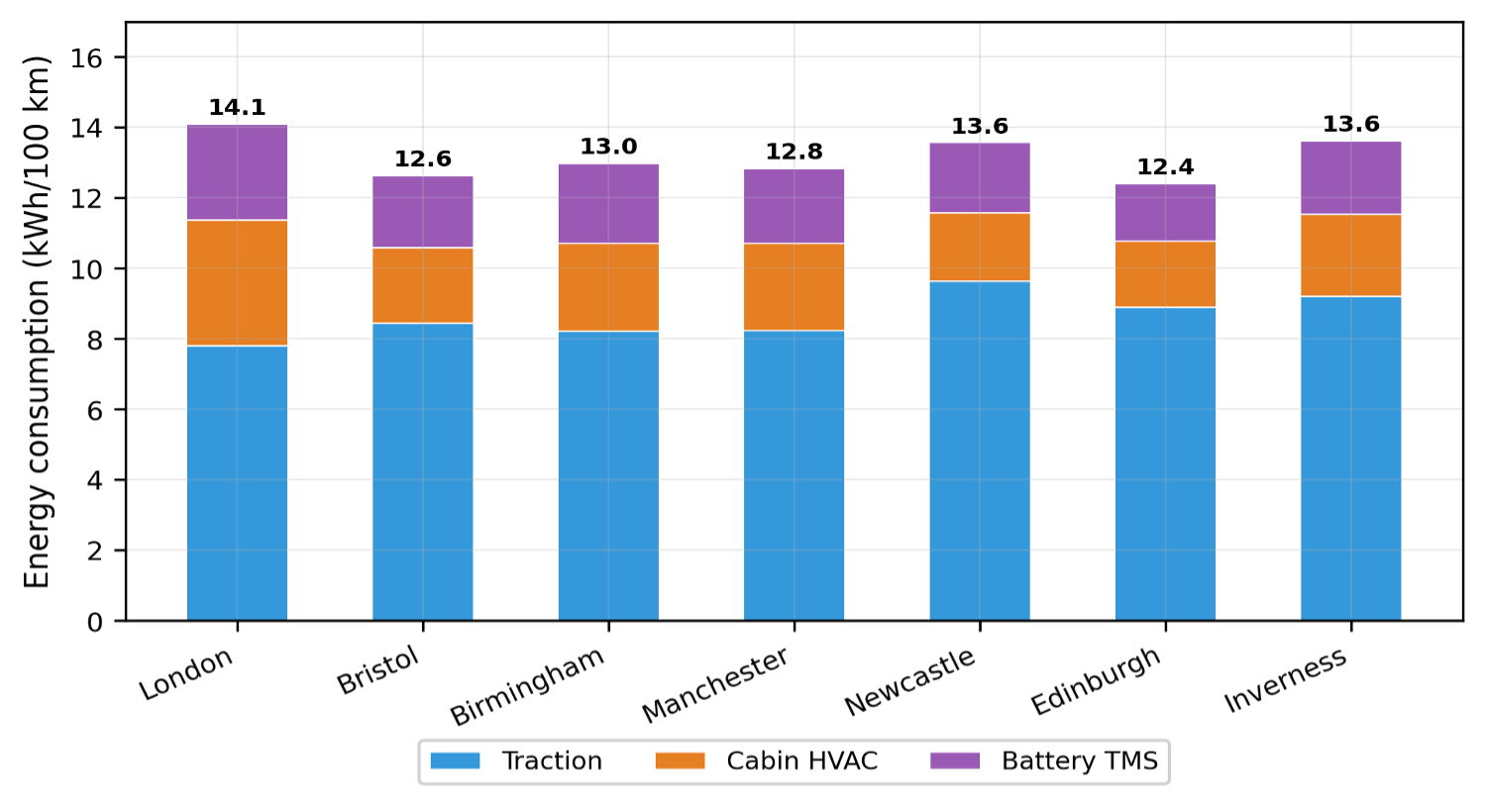}
  \caption{Energy consumption breakdown by city.}
  \label{fig:stacked}
\end{figure}

Across seven UK city loops simulated under identical conditions (100~km, 08:00 departure, winter), total energy consumption ranges from 12.40~kWh/100\,km (Edinburgh) to 14.09~kWh/100\,km (London), a 14\% variation. The HVAC component, however, varies by up to 89\%, from 1.89~kWh/100\,km (Edinburgh) to 3.57 (London), identifying cabin thermal management as the primary differentiator between routes under the winter conditions examined (Table~\ref{tab:summary}, Fig.~\ref{fig:stacked}). Traction remains the largest single component (7.79--9.63~kWh/100\,km), but its inter-city variation is narrower and driven primarily by speed-dependent aerodynamic drag. The combined thermal management load (HVAC + BTMS) accounts for 28--45\% of total energy, underscoring its dominant role in winter driving.

\begin{table}
    \centering
    \caption{Simulation results for seven UK city loops (kWh/100km)  ($\sim$100~km, 5 Jan, 08:00 UTC departure).}
    \label{tab:summary}
    \small
    \begin{tabular*}{\textwidth}{@{\extracolsep{\fill}}lccccccc@{}}
    \hline
    City & Total & Traction  & HVAC  & BTMS  & Thermal (\%) & Speed (kph) & $\bar{T}_\mathrm{amb}$ ($^\circ$C) \\
    London     & 14.09 & 7.79 & 3.57 & 2.73 & 44.7 & 25.3 & 5.42 \\
    Bristol    & 12.62 & 8.43 & 2.15 & 2.05 & 33.2 & 41.2 & 7.38 \\
    Birmingham & 12.98 & 8.20 & 2.50 & 2.28 & 36.8 & 36.2 & 6.40 \\
    Manchester & 12.84 & 8.22 & 2.48 & 2.13 & 36.0 & 35.6 & 6.64 \\
    Newcastle  & 13.56 & 9.63 & 1.93 & 2.00 & 29.0 & 53.0 & 6.83 \\
    Edinburgh  & 12.40 & 8.88 & 1.89 & 1.63 & 28.4 & 37.2 & 8.24 \\
    Inverness  & 13.62 & 9.20 & 2.32 & 2.09 & 32.4 & 45.2 & 5.68 \\
    \hline
    \end{tabular*}
\end{table}

\begin{figure}[!htbp]
  \centering
  \includegraphics[width=0.85\textwidth]{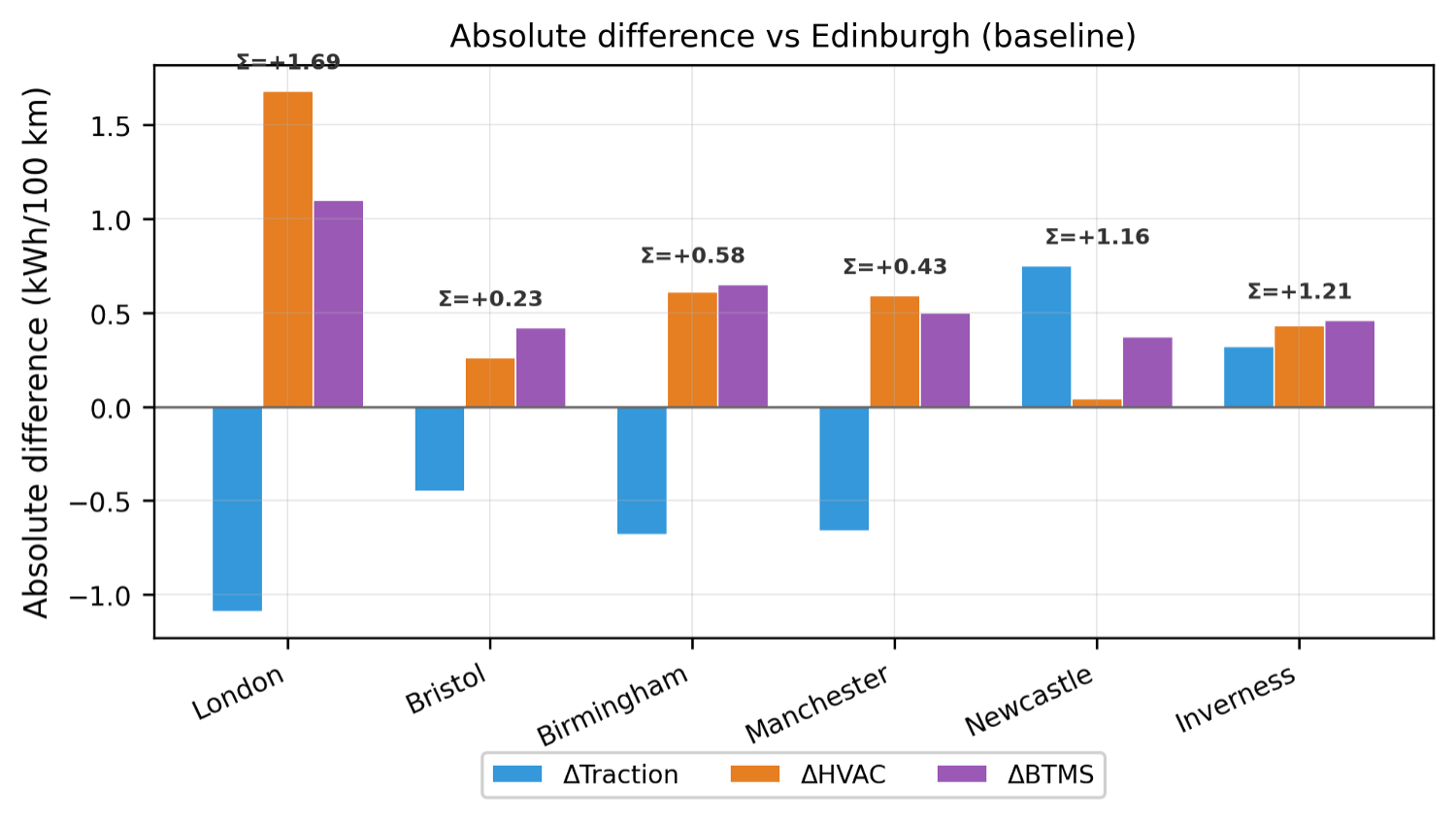}
  \caption{Absolute energy difference by component versus Edinburgh (baseline)}
  \label{fig:abs_diff}
\end{figure}

Decomposing each city's energy relative to Edinburgh (the most efficient) reveals a counterintuitive pattern (Fig.~\ref{fig:abs_diff}). Slow cities (London, Birmingham, Manchester) actually \textit{save} traction energy relative to Edinburgh due to reduced aerodynamic drag at lower speeds, yet their total consumption is \textit{higher} because HVAC and BTMS increases more than offset the traction saving. London is the most extreme case: it saves 1.09~kWh/100\,km in traction but loses 1.68 in HVAC and 1.10 in BTMS, for a net penalty of +1.69~kWh/100\,km. For fast cities (Newcastle, Inverness), the pattern inverts: traction increases dominate, while HVAC differences are small. Newcastle's HVAC is only 0.04~kWh/100\,km above Edinburgh's despite being 1.4~\textdegree C colder, because its high speed shortens the trip sufficiently to compensate.

\begin{figure}[!htbp]
  \centering
  \includegraphics[width=0.75\textwidth]{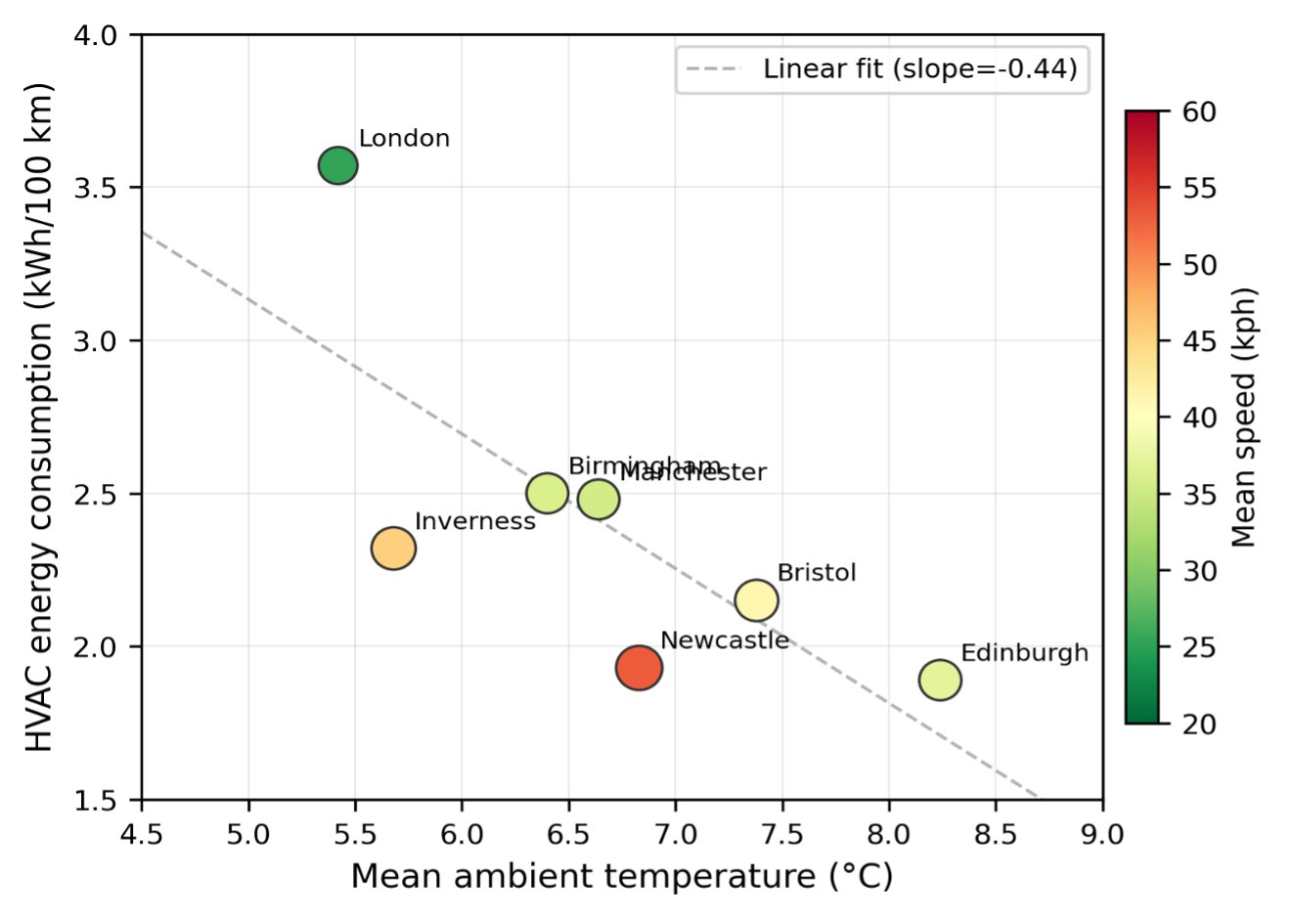}
  \caption{HVAC energy versus ambient temperature. Marker colour and size encode mean driving speed.}
  \label{fig:hvac_tamb}
\end{figure}

The role of speed becomes clearest when HVAC intensity is plotted against ambient temperature with speed encoded as marker colour (Fig.~\ref{fig:hvac_tamb}). A negative correlation is evident, but London deviates dramatically above the trend. At 25.3~kph, London's 100~km loop takes approximately four hours, during which the HVAC must continuously compensate for cabin heat losses. Newcastle, despite a comparable ambient temperature (6.8~\textdegree C), completes the loop in under two hours at 53~kph, yielding 1.93~kWh/100\,km versus London's 3.57, an 85\% difference. HVAC energy per unit distance depends as much on speed as on ambient temperature.

\begin{figure}[!htbp]
  \centering
  \includegraphics[width=0.75\textwidth]{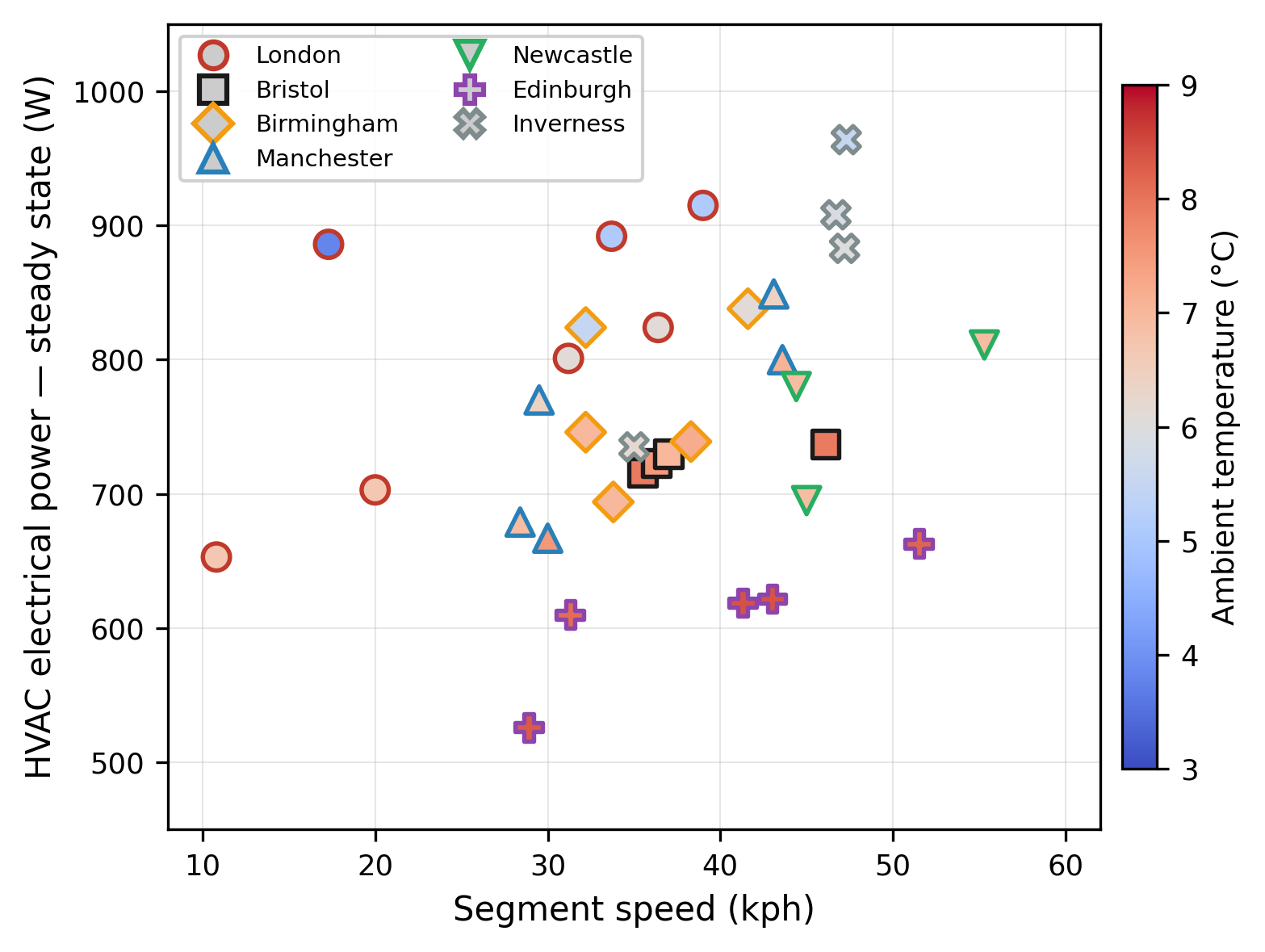}
  \caption{Steady-state HVAC power versus segment speed (cold-start step excluded). Fill colour indicates ambient temperature; edge colour identifies the city.}
  \label{fig:hvac_power_speed}
\end{figure}

At the segment level, two effects interact to determine instantaneous HVAC power (Fig.~\ref{fig:hvac_power_speed}). Higher driving speed increases the cabin-to-ambient thermal conductance through enhanced forced convection, raising instantaneous power demand; within London alone, HVAC power increases 40\% across the 11--39~kph speed range at nearly constant ambient temperature. Simultaneously, colder ambient conditions widen the thermal gradient: Inverness segments near 47~kph at 5.5--6.0~\textdegree C draw 880--960~W, whereas Edinburgh segments at similar speeds but 2.5~\textdegree C warmer require only 620--660~W, roughly 35\% less. Crucially, these two effects compound multiplicatively rather than additively: the 1.83$\times$ ratio between the most and least demanding segments (964~W versus 526~W) far exceeds what either factor alone would produce. The worst-case instantaneous thermal penalty arises when cold ambient and high driving speed coincide.

\begin{figure}[!htbp]
  \centering
  \includegraphics[width=0.85\textwidth]{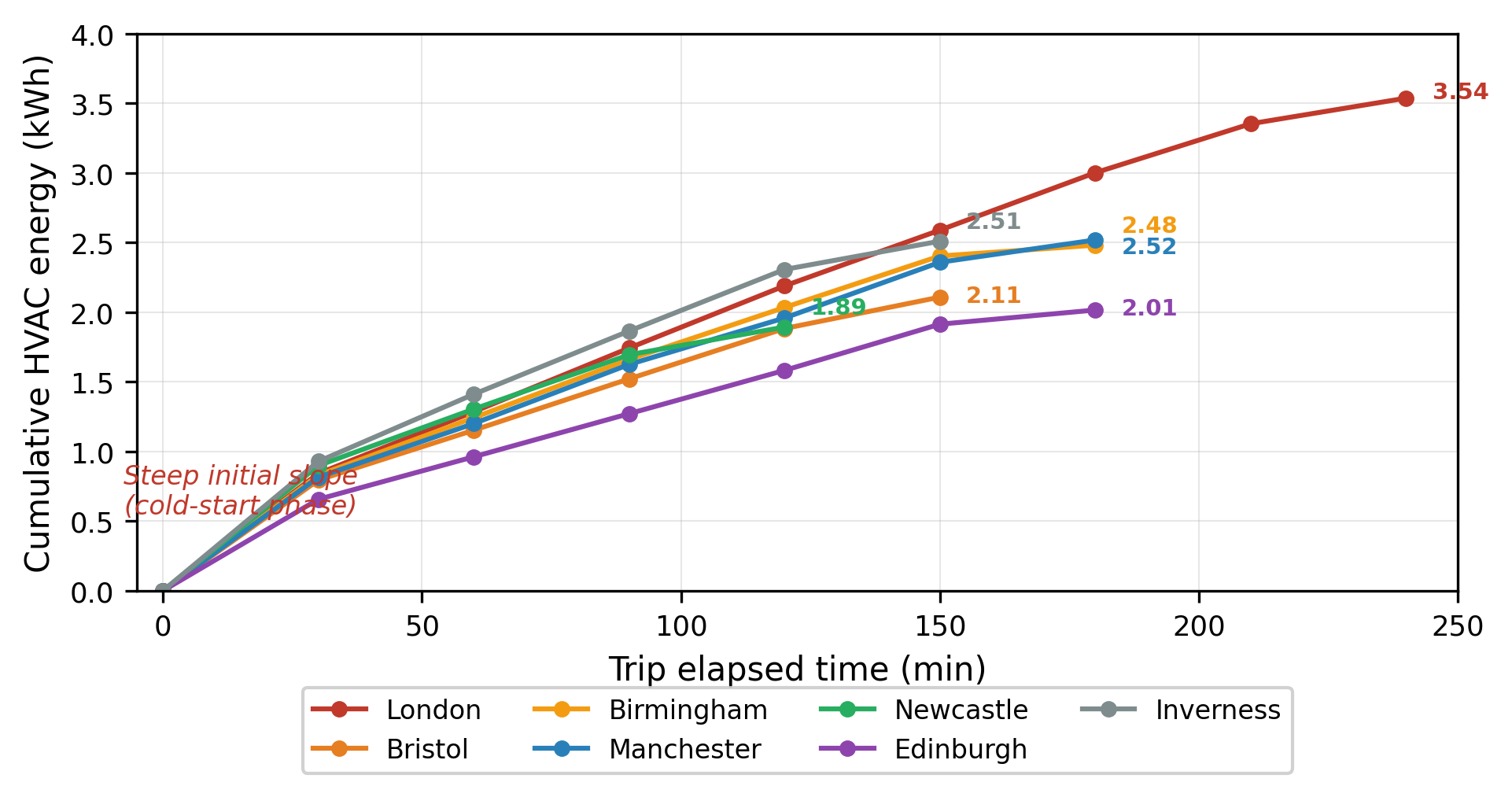}
  \caption{Cumulative HVAC energy consumption over trip duration}
  \label{fig:hvac_timeline}
\end{figure}

The temporal dimension of HVAC energy is best visualised through cumulative energy trajectories (Fig.~\ref{fig:hvac_timeline}). All cities exhibit a steep cold-start phase during the first 30~minutes, during which the cabin is heated from ambient to 22~\textdegree C at peak power of 1300--1850~W. This transient accounts for 25--40\% of the total HVAC energy over the 100~km trip, implying that for shorter commutes ($\leq$30~km) the cold-start penalty alone could exceed half of the thermal energy budget. After the cold-start phase, London's cumulative curve diverges sharply from the other cities, reaching 3.54~kWh over its 240-minute trip, 76\% more than Edinburgh's 2.01~kWh, because its low average speed extends the steady-state integration time far beyond any other route.

\begin{figure}[!htbp]
  \centering
  \includegraphics[width=0.85\textwidth]{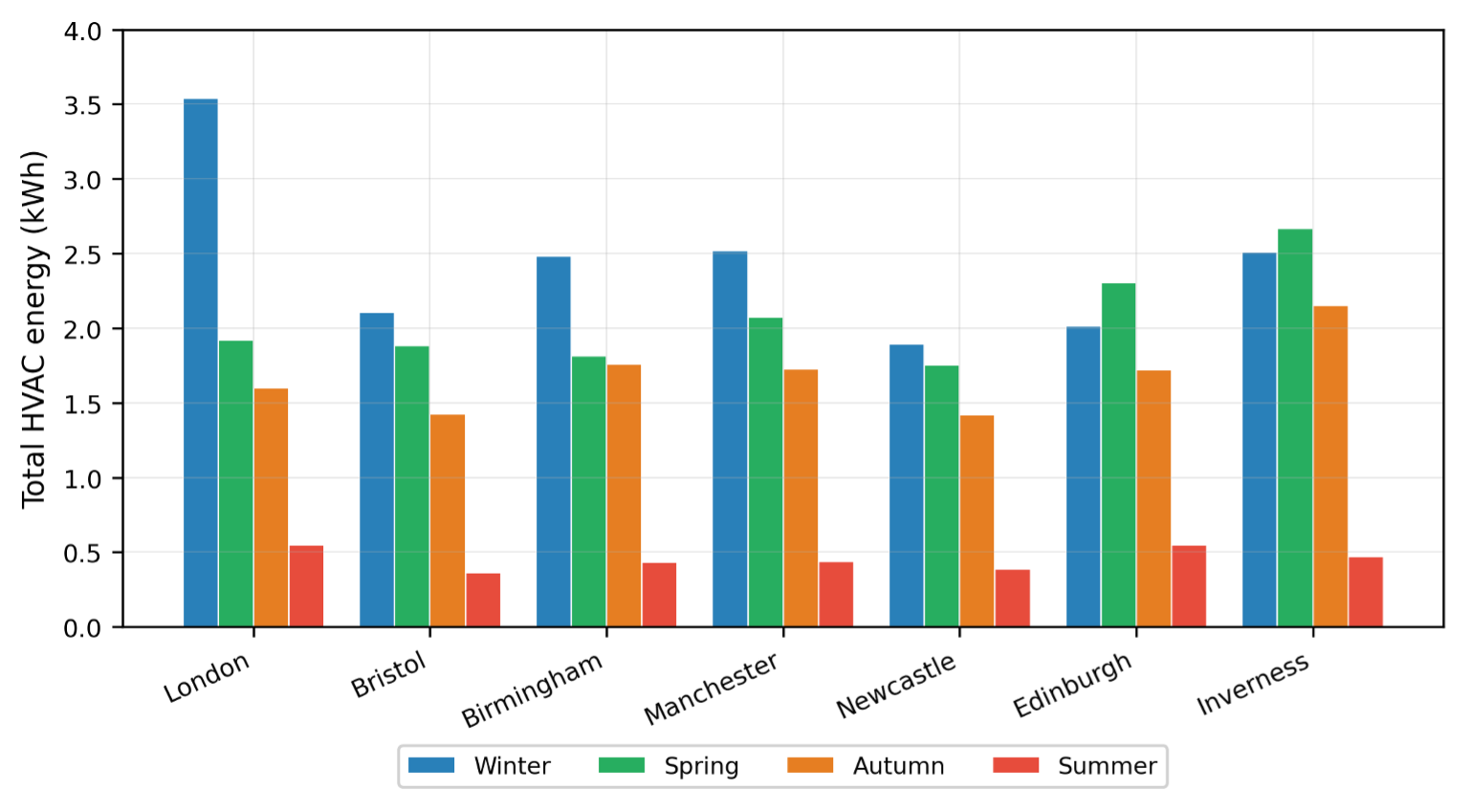}
  \caption{Total HVAC energy by city and season}
  \label{fig:seasonal_bar}
\end{figure}

\begin{figure}[!htbp]
  \centering
  \includegraphics[width=0.75\textwidth]{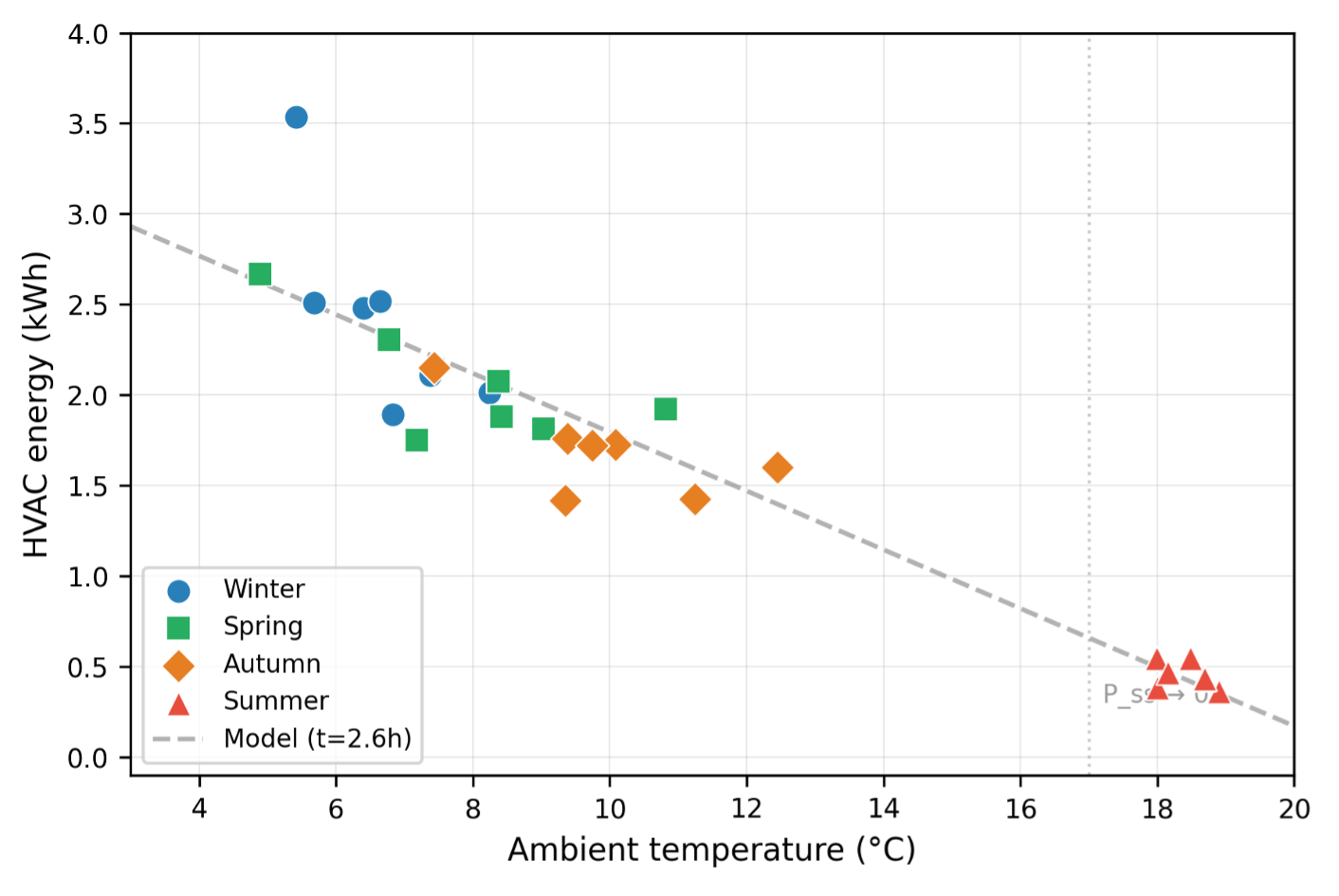}
  \caption{HVAC energy versus ambient temperature for all 28 data points (4 seasons $\times$ 7 cities)}
  \label{fig:hvac_vs_temp_4s}
\end{figure}

These patterns persist across seasons within the UK temperature range examined. Extending the simulation to spring, summer, and autumn (Fig.~\ref{fig:seasonal_bar}) shows that HVAC energy decreases from a winter mean of 2.44~kWh to just 0.45~kWh in summer, an 82\% reduction that follows a smooth, monotonic temperature curve approaching zero near 17~\textdegree C (Fig.~\ref{fig:hvac_vs_temp_4s}). The winter HVAC penalty is not an anomaly but sits on a physically consistent continuum that validates the model's thermal basis. Importantly, the inter-city ranking is not fixed across seasons: Inverness has the highest HVAC in spring (2.67~kWh, coldest at 4.9~\textdegree C), but London has the highest in winter (3.54~kWh, slowest despite not being the coldest). This confirms that the relative importance of temperature versus trip duration shifts with the overall temperature level. At any given temperature, London consistently sits above the trend due to its low speed, while Newcastle sits below, a pattern visible across all four seasons.

The cross-city comparison indicates that both temperature and trip duration drive HVAC energy, but the two factors co-vary across cities, making their individual contributions difficult to isolate. The radial experiment addresses this limitation by holding temperature near-constant while varying only road-network characteristics.

\subsection{Radial routes from Manchester: isolating the road-network effect}
\label{sec:radial}

\begin{figure}[!htbp]
  \centering
  \includegraphics[width=0.85\textwidth]{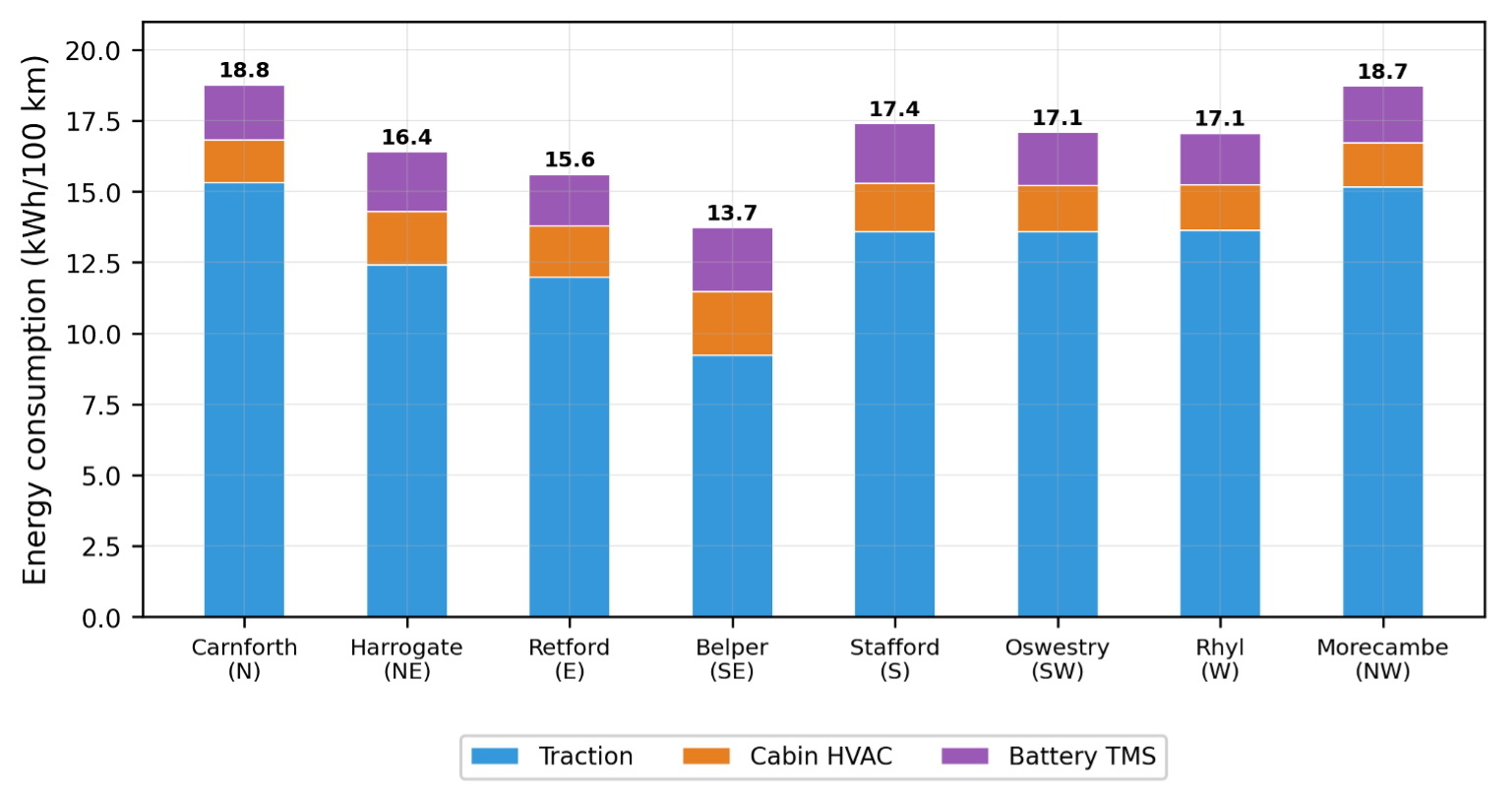}
  \caption{Energy consumption breakdown by route direction.}
  \label{fig:dir_stacked}
\end{figure}

To separate what the cross-city comparison cannot, eight radial routes depart Manchester in different compass directions under near-constant temperature conditions (5.97--6.81~\textdegree C, range 0.84~\textdegree C). The only variable that changes with direction is road type, and consequently average speed, which ranges from 47.8~kph (Belper, SE, via A-roads) to 81.6~kph (Carnforth, N, via the M6 motorway). Under these conditions, HVAC intensity still varies by 47\%, from 1.52~kWh/100\,km (Carnforth) to 2.23 (Belper), demonstrating that road-network characteristics alone drive substantial HVAC variation even when climate is held constant (Table~\ref{tab:direction}, Fig.~\ref{fig:dir_stacked}).

\begin{table}
    \centering
    \caption{Simulation results for eight directional routes from Manchester (kWh/100km) ($\sim$100~km, 08:00 UTC).}
    \label{tab:direction}
    \small
    \begin{tabular*}{\textwidth}{@{\extracolsep{\fill}}lccccccc@{}}
    \hline
    Destination          & Total & Trac.  & HVAC  & BTMS  & Thermal (\%) & Speed (kph) & $\bar{T}_\mathrm{amb}$ ($^\circ$C) \\
    Carnforth (N)  & 18.76 & 15.30 & 1.52 & 1.94 & 18.4 & 81.6 & 6.81 \\
    Harrogate (NE) & 16.40 & 12.41 & 1.88 & 2.11 & 24.3 & 54.0 & 6.39 \\
    Retford (E)    & 15.60 & 11.98 & 1.81 & 1.81 & 23.2 & 64.0 & 6.10 \\
    Belper (SE)    & 13.73 & 9.23  & 2.23 & 2.27 & 32.8 & 47.8 & 5.97 \\
    Stafford (S)   & 17.40 & 13.57 & 1.72 & 2.10 & 22.0 & 69.6 & 6.21 \\
    Oswestry (SW)  & 17.09 & 13.59 & 1.63 & 1.87 & 20.5 & 73.9 & 6.28 \\
    Rhyl (W)       & 17.05 & 13.63 & 1.61 & 1.81 & 20.1 & 75.0 & 6.28 \\
    Morecambe (NW) & 18.72 & 15.15 & 1.56 & 2.01 & 19.1 & 76.4 & 6.81 \\
    \hline
    \end{tabular*}
\end{table}

\begin{figure}[!htbp]
  \centering
  \includegraphics[width=0.6\textwidth]{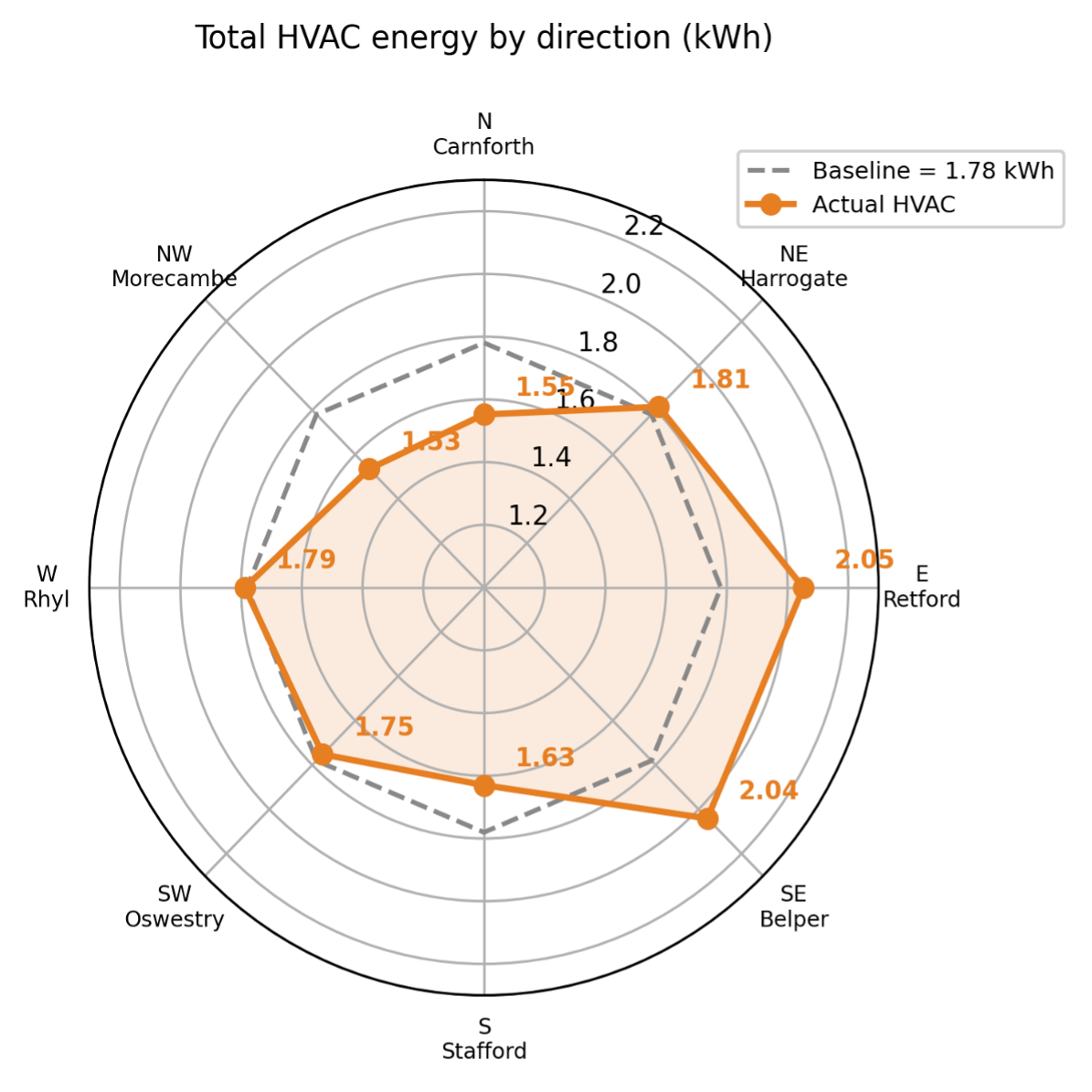}
  \caption{Polar plot of total HVAC energy (kWh) by route direction.}
  \label{fig:dir_polar}
\end{figure}

\begin{figure}[!htbp]
  \centering
  \includegraphics[width=\textwidth]{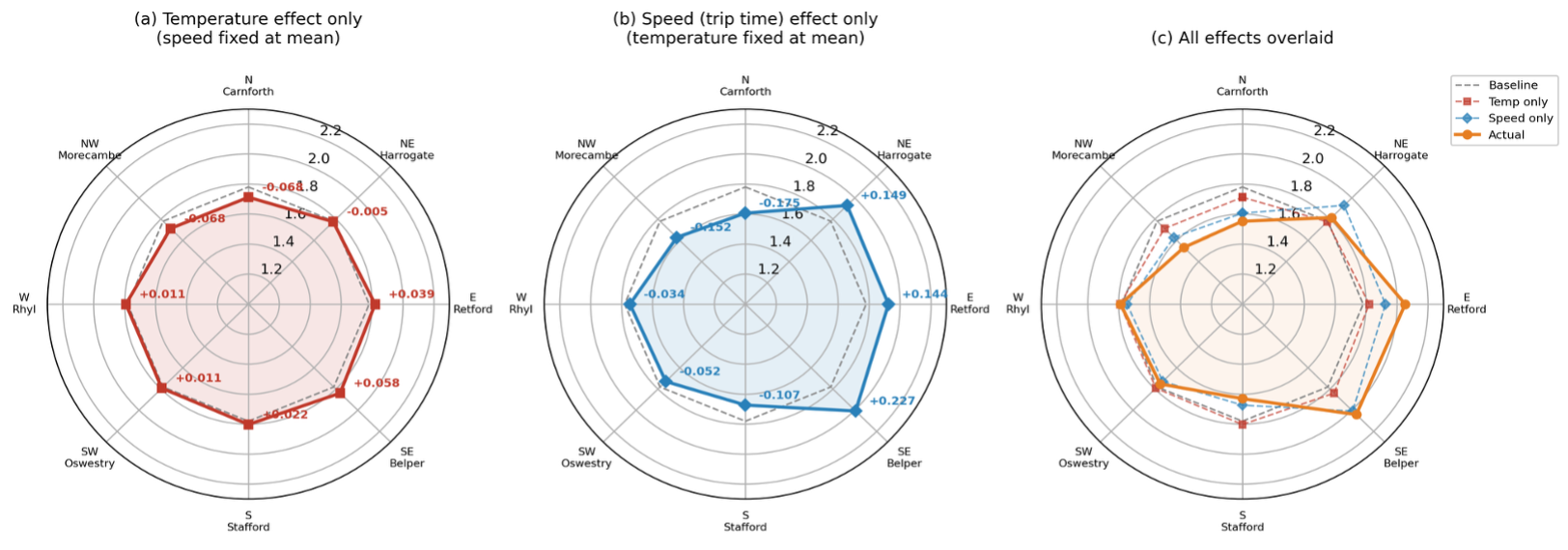}
  \caption{Decomposition of the polar shape distortion}
  \label{fig:dir_decomp}
\end{figure}

The directional pattern of HVAC energy is asymmetric: the polar plot (Fig.~\ref{fig:dir_polar}) is elongated toward E/SE/NE and compressed toward N/NW, reflecting directional differences in road type and trip time. Decomposing this shape into temperature-only and speed-only effects (Fig.~\ref{fig:dir_decomp}) reveals that the temperature-only shape is nearly circular (distortion range 0.13~kWh), while the speed-only shape closely reproduces the actual polar pattern (distortion range 3.2$\times$ larger). The E/SE/NE directions from Manchester pass through the Peak District and Yorkshire via slower A-roads, extending trip times to 1.8--1.9~hours; the N/NW directions take the M6 motorway at ${>}76$~kph, completing trips in 1.2--1.3~hours. Speed accounts for 92\% of the polar shape distortion; temperature contributes less than 8\%. The polar asymmetry is a road-network effect, not a climate effect.

\begin{figure}[!htbp]
  \centering
  \includegraphics[width=0.85\textwidth]{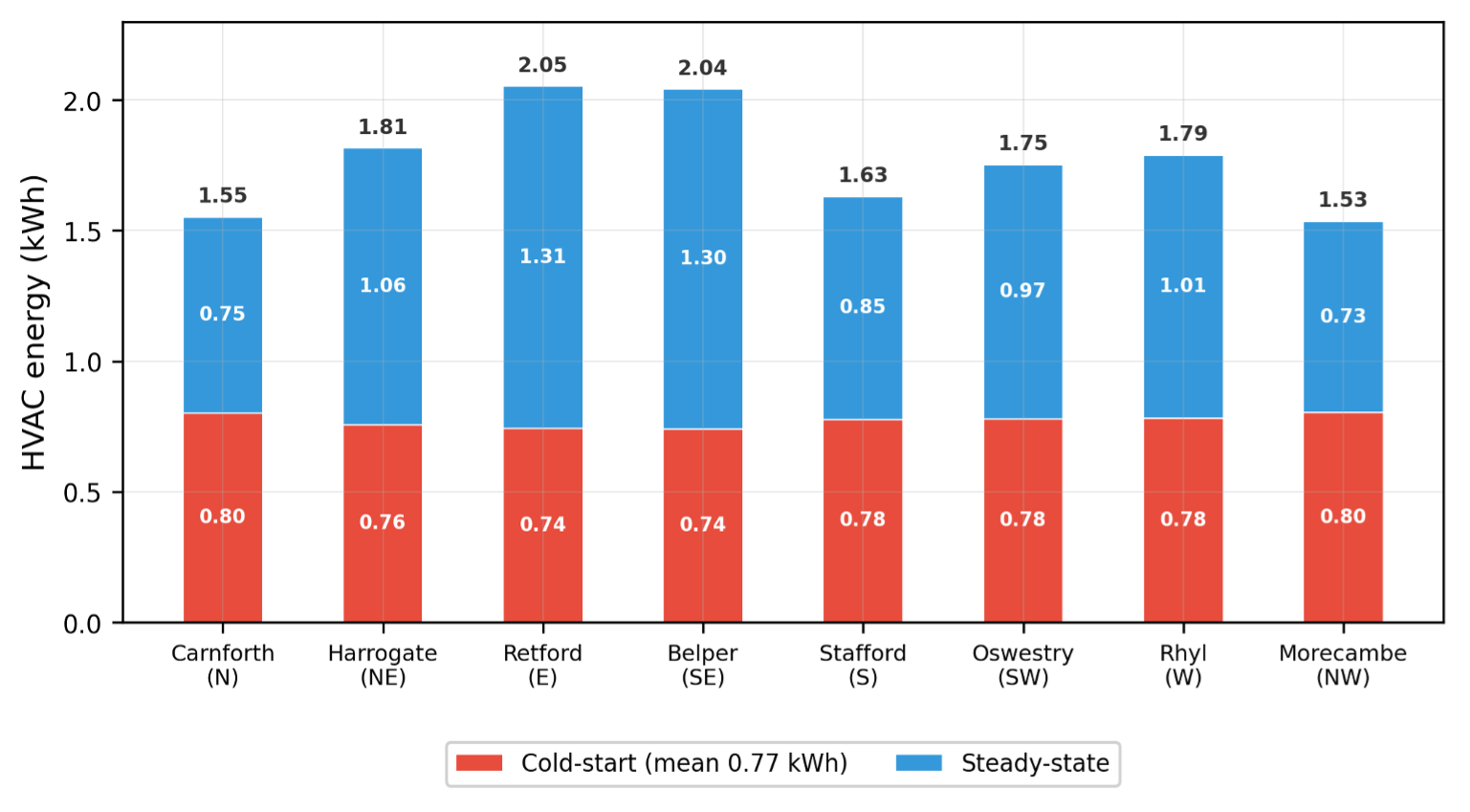}
  \caption{HVAC energy decomposition: cold-start (red, $\sim$0.77~kWh constant) versus steady-state (blue, varies with trip duration)}
  \label{fig:dir_cs_ss}
\end{figure}

Separating cold-start from steady-state contributions (Fig.~\ref{fig:dir_cs_ss}) indicates that, under the near-constant temperature conditions of this experiment, inter-route variation originates primarily from how long the heat pump runs rather than from how hard it works. Cold-start energy is effectively constant across all eight directions at $0.77 \pm 0.02$~kWh, depending solely on the initial cabin-to-setpoint temperature differential. Steady-state power is also near-uniform at $850 \pm 8\%$~W. Consequently, routes with slower speeds (Retford at 64~kph, Belper at 48~kph) accumulate more steady-state energy not because their thermal conditions are more demanding, but because the heat pump runs longer.

Both the cross-city and radial experiments suggest that temperature and trip duration are the principal drivers of HVAC energy variation, but their relative importance varies by route context. To quantify each factor's independent contribution and derive a general prediction model, the following section pools all 15 routes from both experiments into a unified regression.

\subsection{Decomposition of HVAC energy: temperature versus trip duration}
\label{sec:decomposition_synthesis}

In the cross-city experiment, temperature and speed co-vary; in the radial experiment, temperature is held nearly constant. Combining both datasets into a single regression allows separation of the two factors. Because speed affects HVAC energy through trip duration (i.e.\ how long the heat pump must run), the regression uses trip time $t = D/v$ (hours) rather than speed directly. The model takes the form:
\begin{equation}
  E_\mathrm{hvac} = c_0 + c_T \cdot T_\mathrm{amb} \cdot t + c_t \cdot t
  \label{eq:model}
\end{equation}
where $c_0$ represents a fixed cold-start overhead, $c_T \cdot T \cdot t$ captures the temperature--time interaction (warmer ambient reduces the loss rate, and this saving scales with trip duration), and $c_t \cdot t$ represents the base steady-state energy accumulation independent of temperature.

\begin{figure}[!htbp]
  \centering
  \includegraphics[width=0.6\textwidth]{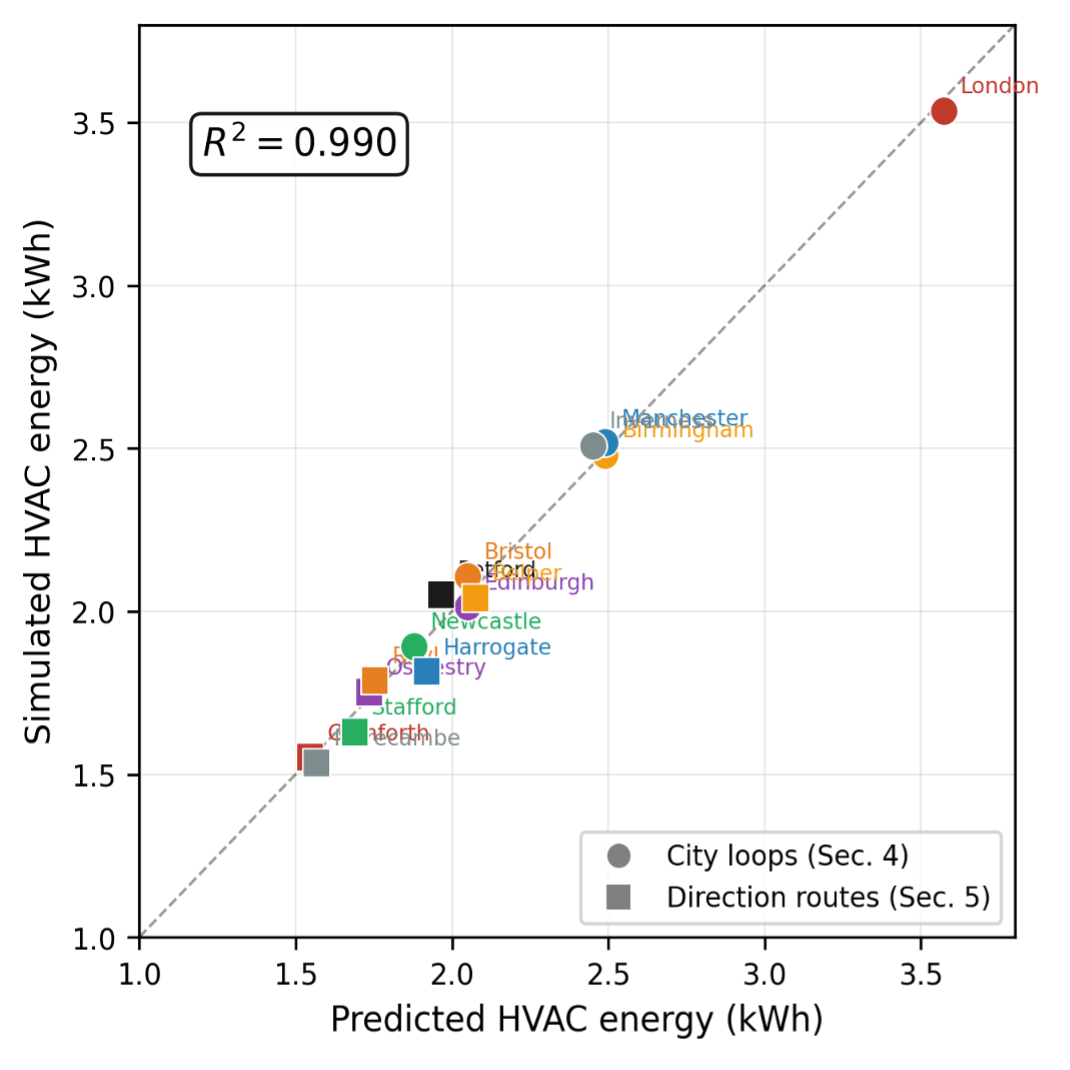}
  \caption{Predicted versus simulated HVAC energy for all 15 routes}
  \label{fig:pred_actual}
\end{figure}

The fitted model across all 15 routes is:
\begin{equation}
  E_\mathrm{hvac} = 0.847 + (-0.098) \times T_\mathrm{amb} \times t + 1.230 \times t \qquad (R^2 = 0.990)
  \label{eq:fitted}
\end{equation}
\label{sec:decomp_results}
The three coefficients have direct physical interpretations. The cold-start overhead $c_0 = 0.847$~kWh is a fixed energy cost per trip, independent of route length, speed, or temperature, and is consistent with the independently measured cold-start values from both experiments (0.77--0.85~kWh). The temperature coefficient $c_T = -0.098$~kWh/(\textdegree C$\cdot$h) captures the thermodynamic benefit of warmer ambient: each 1~\textdegree C increase saves 0.098~kWh per hour of driving by reducing the cabin-to-ambient thermal gradient. The base rate $c_t = 1.230$~kWh/h represents HVAC energy accumulation at a hypothetical 0~\textdegree C ambient; combined with $c_T$, the effective steady-state HVAC power is $P_\mathrm{ss} = (1230 - 98 \times T_\mathrm{amb})$~W, giving approximately 593~W at the dataset mean of 6.5~\textdegree C. All 15 data points cluster tightly around the 1:1 prediction line with no systematic bias (Fig.~\ref{fig:pred_actual}), confirming that a single two-variable model generalises across different origins, directions, and distance ranges. Because the decomposition depends only on the regression structure and mean-deviation expansion, the same methodology can be applied to other vehicles, climates, and route networks by re-fitting the three coefficients to local data.

\begin{figure}[!htbp]
  \centering
  \includegraphics[width=0.85\textwidth]{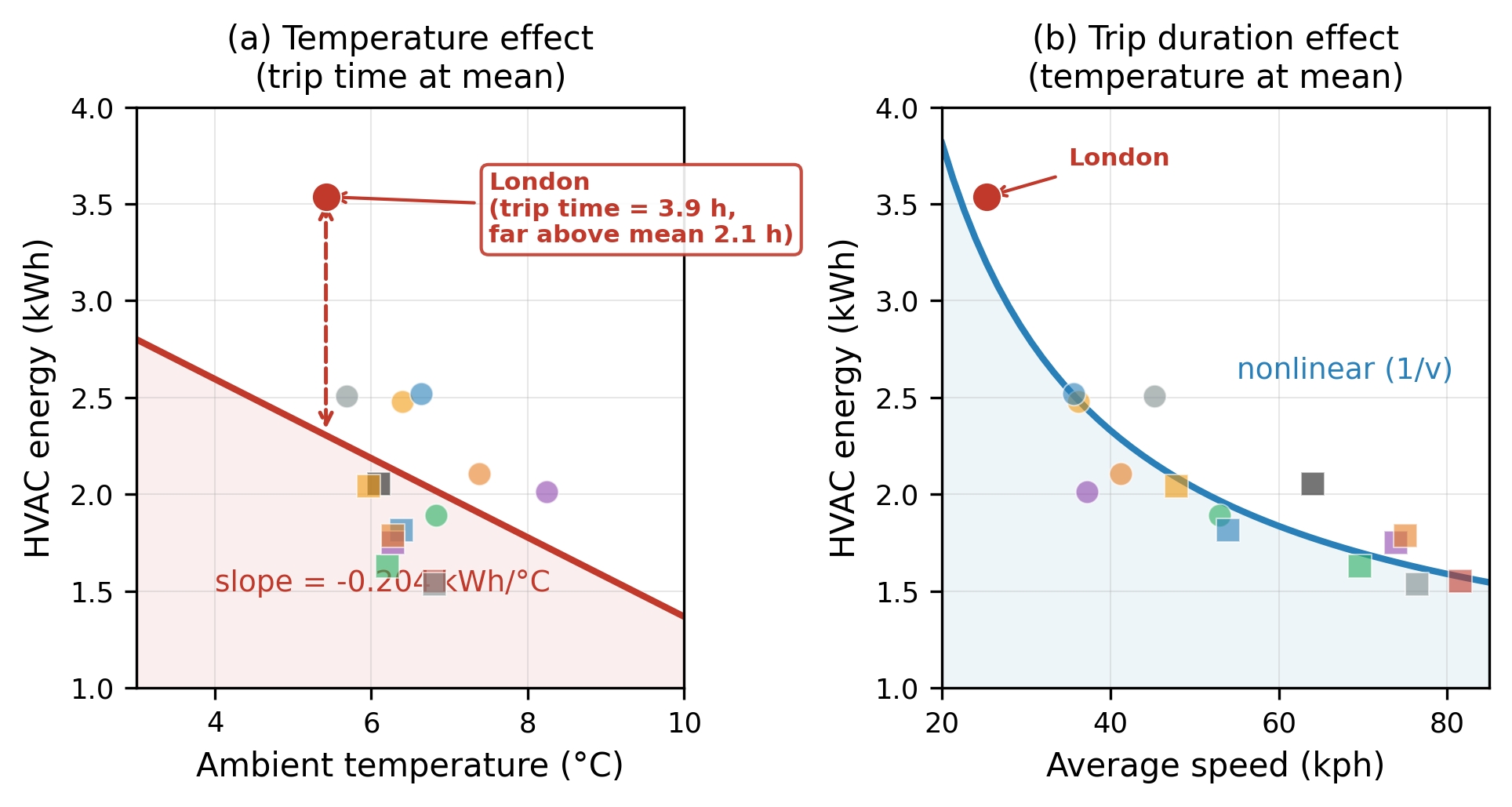}
  \caption{Sensitivity analysis. (a) HVAC versus ambient temperature at constant mean trip time. (b) HVAC versus speed at constant mean temperature. The $1/v$ curvature in (b) reflects the nonlinear trip duration effect.}
  \label{fig:sensitivity}
\end{figure}

Sweeping each variable while holding the other at its dataset mean ($\bar{t} = 2.09$~h; $\bar{T} = 6.5$~\textdegree C) reveals a fundamental asymmetry between the two predictors (Fig.~\ref{fig:sensitivity}). Temperature alone is a poor predictor: data points scatter widely around the temperature-only line (residual standard deviation $= 0.52$~kWh), and London sits 1.23~kWh above the prediction because its extreme trip time (3.91~h, nearly double the mean) cannot be captured by a temperature-only view. Of London's total deviation from the mean (+1.52~kWh), only approximately 0.3~kWh is attributable to its below-average temperature (5.42~\textdegree C); the remaining 1.2~kWh is entirely a trip-duration effect. Trip duration alone is far more explanatory (residual standard deviation $= 0.27$~kWh), with a strongly nonlinear $1/v$ dependence: below 40~kph, HVAC energy is extremely sensitive to speed (a 10~kph reduction from 40 to 30~kph increases HVAC by 0.61~kWh, or 36\%), while above 60~kph the returns from further speed increases diminish rapidly (70 to 80~kph saves only 0.10~kWh). London falls squarely in the high-sensitivity regime, confirming that its high HVAC is accurately captured by the speed model. This asymmetry illustrates the central thesis of this paper: \textit{ambient temperature alone is an insufficient predictor of winter HVAC energy; trip duration must be considered as an independent variable.}

Because the model contains a temperature--time interaction term, the contribution of each factor to a given route's HVAC energy cannot be read directly from the coefficients. Instead, each route's temperature and trip time are expressed as deviations from the dataset means ($\bar{T} = 6.50$~\textdegree C, $\bar{t} = 2.09$~h), and the interaction term is expanded:
\begin{equation}
  T_i \times t_i = (\bar{T} + \Delta T)(\bar{t} + \Delta t) = \bar{T}\bar{t} + \bar{t}\cdot\Delta T + \bar{T}\cdot\Delta t + \Delta T \cdot \Delta t
  \label{eq:expand}
\end{equation}
Substituting into the model and grouping terms, each route's deviation from the mean HVAC ($\bar{E} = 2.08$~kWh) decomposes into a temperature contribution $c_T \times \bar{t} \times \Delta T_i$ (the effect of the route being warmer or colder than average, evaluated at mean trip time), a trip-time contribution $(c_T \times \bar{T} + c_t) \times \Delta t_i$ (the effect of the route being faster or slower, evaluated at mean temperature), and a small residual capturing the interaction and model error. The temperature share for each route is $|\Delta E_\mathrm{temp}| / (|\Delta E_\mathrm{temp}| + |\Delta E_\mathrm{time}|)$.

\begin{figure}[!htbp]
  \centering
  \includegraphics[width=0.85\textwidth]{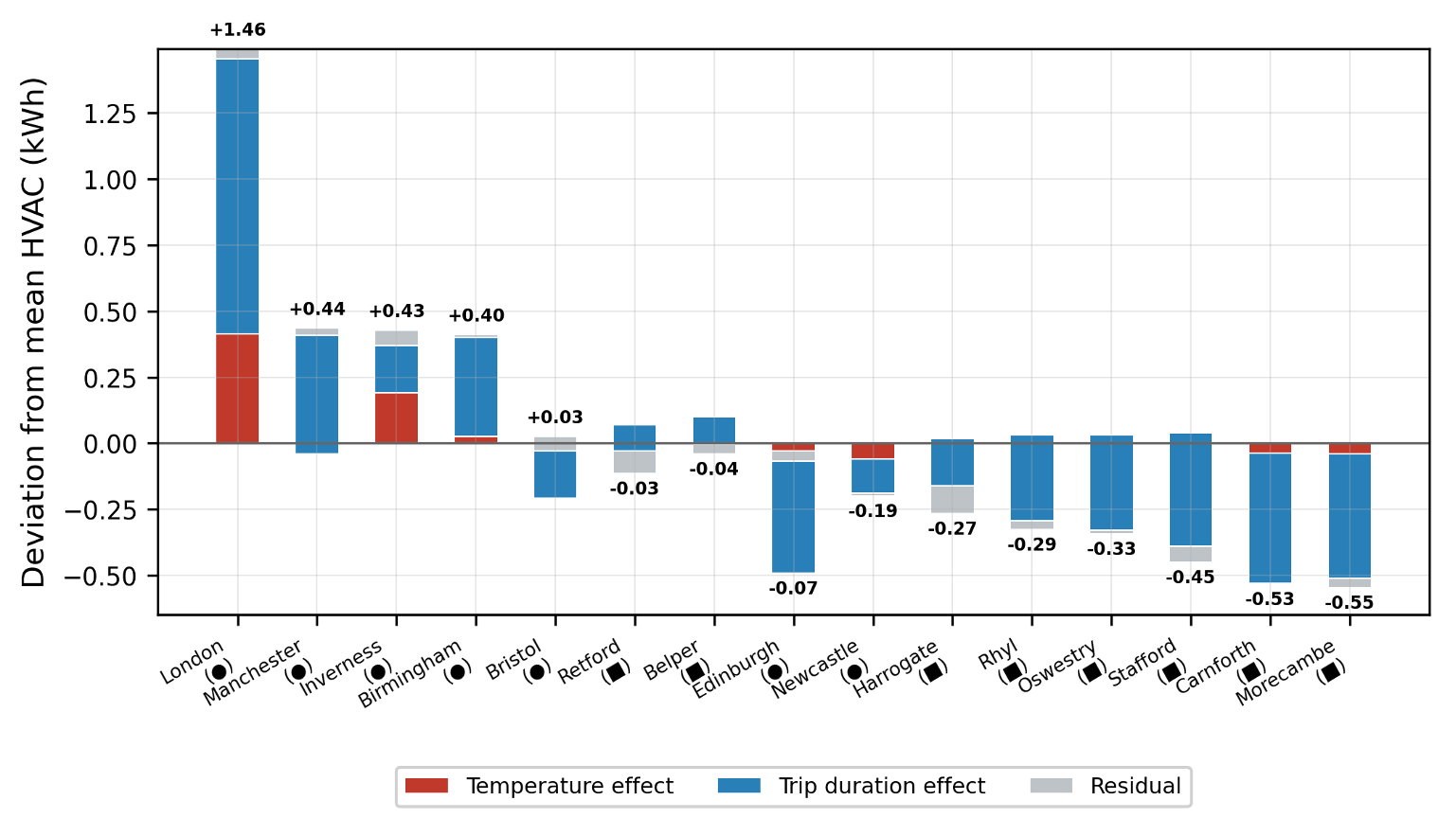}
  \caption{Attribution of HVAC deviation from the mean}
  \label{fig:attribution}
\end{figure}

A worked example illustrates the method. London has $T = 5.42$~\textdegree C ($\Delta T = -1.08$~\textdegree C) and $t = 3.91$~h ($\Delta t = +1.83$~h):
\begin{align}
  \text{Temperature contribution} &= (-0.098) \times 2.09 \times (-1.08) = +0.22~\text{kWh} \notag \\
  \text{Trip time contribution} &= ((-0.098) \times 6.50 + 1.230) \times (+1.83) = +1.08~\text{kWh} \notag \\
  \text{Temperature share} &= |0.22| \,/\, (|0.22| + |1.08|) = 17\% \notag
\end{align}
London's HVAC is 1.46~kWh above the mean; of this, only 17\% comes from being colder than average, while 83\% comes from its extreme trip time. Within the model's two-factor framework, London's high HVAC is predominantly a congestion problem rather than a temperature problem. Applying this decomposition to all 15 routes (Fig.~\ref{fig:attribution}) reveals three natural categories. Most routes are trip-time-dominated (temperature share ${<}20\%$): London (17\%), Birmingham (5\%), Manchester (6\%), and all motorway-accessible radial routes (Carnforth, Stafford, Oswestry, Rhyl, Morecambe, all 11--12\%), whose HVAC deviations are almost entirely explained by whether they are slower or faster than average. A smaller group is temperature-influenced (share ${>}40\%$): Bristol (50\%), Inverness (48\%), and Belper (51\%), where temperatures deviate noticeably from the mean while trip times are closer to average. Edinburgh represents a striking case of near-cancellation: its warm temperature ($\Delta T = +1.74$~\textdegree C) saves 0.36~kWh, but its moderate speed extends the trip ($\Delta t = +0.78$~h), adding 0.46~kWh. The two effects nearly cancel, leaving Edinburgh almost exactly at the overall mean despite having the warmest ambient of all city loop routes, a coincidence that would be invisible without the decomposition.

\begin{figure}[!htbp]
  \centering
  \includegraphics[width=0.85\textwidth]{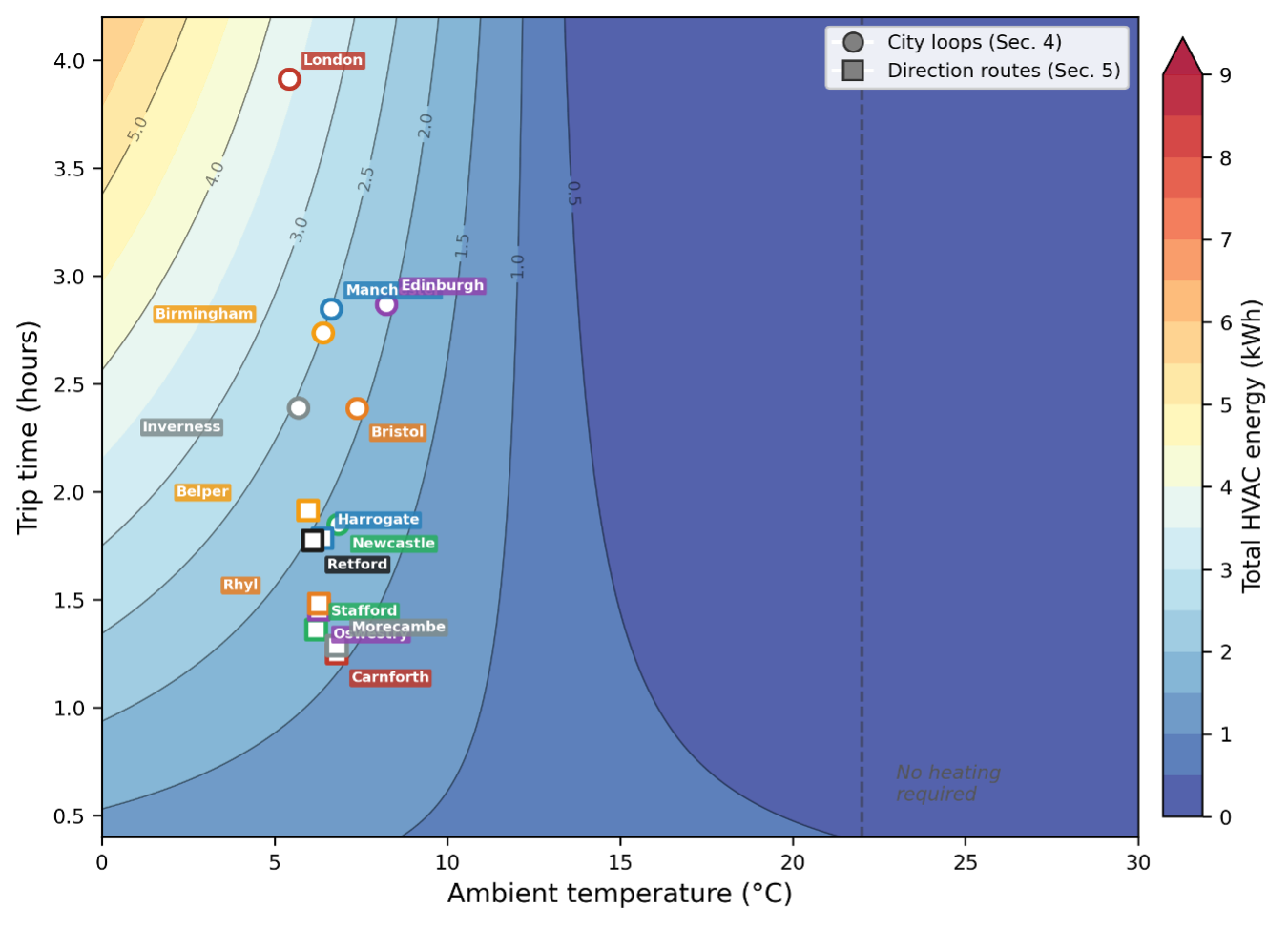}
  \caption{HVAC energy landscape as a function of ambient temperature and trip time. Full temperature range (0--30~\textdegree C) shown; dashed line marks $T_\mathrm{set} = 22$~\textdegree C beyond which no heating is required. All 15 routes from both experiments overlaid.}
  \label{fig:contour}
\end{figure}

Placing all 15 routes on a single temperature $\times$ trip-time contour map (Fig.~\ref{fig:contour}) provides a unified view of the HVAC energy landscape. Contour lines compress at long trip times, confirming that slow driving causes disproportionate HVAC penalties. The radial routes cluster in a horizontal band near 1--2~hours at 6--7~\textdegree C, while the city loop routes span a wider vertical range up to 4~hours, with London isolated in the high-energy corner. Above approximately 15~\textdegree C, HVAC energy drops below 1~kWh even for long trips, and above the 22~\textdegree C cabin setpoint the heating demand vanishes entirely.

The decomposition yields a parsimonious three-input prediction model for HVAC energy:
\begin{equation}
  E_\mathrm{hvac} = 0.85 + (1.23 - 0.10 \times T_\mathrm{amb}) \times \frac{D}{v} \qquad \text{[kWh]}
  \label{eq:practical}
\end{equation}
The cold-start term (0.85~kWh) is a fixed per-trip overhead; for short commutes ($\leq$30~km at 40~kph), this alone exceeds 40\% of total HVAC energy, making preconditioning while plugged in the highest-impact intervention. The steady-state power $P_\mathrm{ss} = (1230 - 98 \times T_\mathrm{amb})$~W is temperature-dependent, yielding approximately 740~W at 5~\textdegree C and 446~W at 8~\textdegree C. The trip time $t = D/v$ is the integration window; the $1/v$ nonlinearity means that rerouting from a 25~kph urban road to a 50~kph bypass halves the trip time and thus halves the steady-state HVAC energy. The model requires only three readily available inputs (ambient temperature, average speed, and distance), explains 99.0\% of the HVAC variance across 15 diverse winter routes, and can be directly integrated into route planning systems to provide a physics-grounded HVAC energy correction alongside traction-based range estimates.

\section{Discussion}

\subsection{Interpretation of Key Findings}

The results indicate that, under the conditions examined (UK winter, relatively flat terrain, single vehicle model), real-world EV HVAC energy consumption is shaped primarily by the compound interaction of two externally imposed factors, ambient temperature and traffic-determined driving speed, rather than by either factor alone. This has practical consequences for how range variability is understood and managed: treating HVAC energy as a climate-only phenomenon, as most existing analyses do, risks underestimating the energy penalty in congested urban environments and misdirecting the interventions designed to address it. Three principal insights emerge.

First, temperature and speed affect HVAC energy through distinct and partially opposing mechanisms. Lower ambient temperature increases the instantaneous HVAC power demand by widening the cabin-to-ambient thermal gradient. Higher speed further increases instantaneous power through enhanced forced convection. However, higher speed simultaneously \emph{reduces} HVAC energy per unit distance by shortening the trip duration over which the heat pump must operate. This creates a trade-off that is not captured by studies evaluating temperature or speed in isolation: the worst HVAC \emph{power} occurs under cold, fast conditions, but the worst HVAC \emph{energy per kilometre} occurs under cold, slow conditions where extended trip duration dominates.

Second, the per-route attribution analysis reveals that the relative importance of temperature and trip duration is context-dependent. In the cross-city comparison, London's HVAC energy exceeds the mean by 1.46~kWh, of which 83\% is attributable to its extreme trip duration (3.9~hours at 25.3~kph) and only 17\% to its below-average temperature. By contrast, routes such as Bristol and Inverness show an approximately equal split between temperature and trip-duration contributions. Edinburgh demonstrates near-cancellation: its warm climate saves energy, but its moderate speed extends trip time, leaving it close to the overall mean. These patterns would be invisible without the decomposition.

Third, the practical prediction model (Eq.~\ref{eq:practical}) shows that the combined effect of temperature and trip duration can be captured by a simple three-input formula with $R^2 = 0.99$. The $1/v$ nonlinearity in this model has direct operational implications: below 40~kph, HVAC energy is highly sensitive to speed, while above 60~kph the returns from further speed increases diminish. This suggests that congestion relief in slow urban environments offers disproportionate HVAC energy savings.

\subsection{Implications for Infrastructure and Energy Equity}

The spatially uneven distribution of HVAC energy penalties has implications for charging infrastructure planning. Across the seven UK city loops, HVAC energy intensity varies by up to 89\% (London versus Edinburgh) for identical vehicles and trip distances. Because thermal loads scale with trip duration, cities characterised by chronic traffic congestion and cold winters impose a compound energy penalty on EV users. If a fixed fraction of energy is replenished via public chargers, the required charger density per EV scales with kWh/km, implying that cities like London may require substantially more public charging capacity per registered EV than cities like Edinburgh to maintain equivalent service levels. To illustrate the magnitude: London's 14.09~kWh/100\,km versus Edinburgh's 12.40~kWh/100\,km represents a 14\% higher energy demand per kilometre. For a fleet of 100,000~EVs each commuting 30~km daily, this difference translates to approximately 50~MWh of additional daily energy demand, equivalent to roughly 200 additional 7~kW chargers operating four hours per day.

This finding intersects with growing evidence that EV charging infrastructure is already inequitably distributed. Yu et al.\ \cite{yu2025_ev_charging_equity} found that disadvantaged communities in the United States have 64\% fewer public charging stations per capita than non-disadvantaged areas, and that this disparity has not improved despite a 350\% increase in total charger deployment between 2015 and 2021. If charger allocation decisions are based on vehicle population alone rather than on spatiotemporal energy demand, congested and cold cities will be systematically under-served relative to their actual per-kilometre energy requirements. Bauer et al.\ \cite{bauer2021might} similarly noted that lower-income EV drivers, who are more likely to depend on public charging and to commute through congested urban corridors, may face disproportionate operating costs. The compound thermal-and-congestion penalties documented in this study add a previously unquantified dimension to this energy equity concern.

The decomposition results also suggest that congestion mitigation may be a more effective and durable policy lever than climate adaptation for reducing HVAC energy penalties. The U.S.\ Department of Energy reported that cold-weather range loss reaches 59\% under urban driving conditions (with ample idle time) but only 39\% under highway conditions \cite{DOE2024_cold_BEV}; the 20~percentage-point gap is attributable to extended trip duration, consistent with this study's finding that trip time dominates HVAC energy variability in congested settings. Moreover, as ambient temperatures rise under climate change, heating-related energy penalties will gradually diminish, but congestion-induced trip-time penalties will persist. This asymmetry implies that investments in congestion reduction, through traffic management, modal shift, or road-network design, offer compounding benefits for EV energy efficiency that outlast the current climate conditions.

\subsection{Limitations and Future Work}

Several limitations should be noted. First, the study focuses on a single vehicle model (Tesla Model Y RWD); while the framework is designed for vehicle-agnostic parameterisation, the specific regression coefficients in the practical model are vehicle-dependent and should be re-fitted for other platforms. Second, the UK routes used in the case studies are relatively flat; elevation and gradient effects, which can significantly affect traction energy in mountainous terrain, are not modelled. Third, wind speed and humidity, which affect aerodynamic drag and cooling-mode COP respectively, are not included as dynamic inputs. Fourth, the spatiotemporal temperature database uses reanalysis data at $0.625^\circ \times 0.5^\circ$ resolution, which may not capture micro-climate effects in dense urban environments. Fifth, the HVAC model and the practical prediction formula (Eq. 26) address heating demand only. In the UK climate examined, heating accounts for the dominant share of cabin thermal energy; cooling loads arise only at ambient temperatures above the cabin setpoint and were not observed in the simulation conditions. A cooling-mode formulation would require independently fitted coefficients, as the COP characteristics of air-conditioning differ from those of heat-pump heating, and additional thermal loads such as solar radiation through glazing become significant.

Future work could extend the framework to incorporate elevation profiles and wind data, apply it to international contexts with greater climatic diversity (e.g., Scandinavian versus Mediterranean cities), and validate the practical prediction model against fleet-scale telematics data. The decomposition method is transferable and could be applied to cooling-dominated climates to assess whether the trip-duration dominance observed in heating mode persists under air conditioning loads.

\section{Conclusion}

This study develops a spatiotemporal simulation framework that couples traffic-aware driving speed, location- and time-specific ambient temperature, and physics-based energy submodels to quantify EV energy consumption under realistic conditions. Applied through a factorial experimental design across UK cities and routes, the framework yields three principal contributions.

First, the compound interaction between ambient temperature and traffic-determined trip duration is identified as a principal driver of HVAC energy variability under the conditions studied. Across seven UK city loops, total energy consumption varies by 14\%, but the HVAC component varies by up to 89\%, identifying cabin thermal management as the primary differentiator between routes in winter. The multiplicative interaction between temperature and speed produces a 1.83$\times$ ratio between the most and least demanding driving segments.

Second, a regression-based decomposition method quantifies the per-route contributions of temperature and trip duration to HVAC energy. The analysis reveals that trip duration is frequently the dominant factor: in London, 83\% of the above-average HVAC energy is attributable to congestion-extended trip time rather than temperature. In the radial experiment, where temperature is held near-constant, road-network characteristics account for 92\% of the directional variation in HVAC energy.

Third, the decomposition yields a practical three-input prediction model ($E_\mathrm{hvac} = 0.85 + (1.23 - 0.10 \times T_\mathrm{amb}) \times D/v$) that explains 99.0\% of HVAC energy variance across 15 diverse routes. The model identifies a fixed cold-start overhead of 0.85~kWh per trip and a strongly nonlinear speed sensitivity below 40~kph, providing actionable guidance for route planning and infrastructure design.

Together, these findings suggest that EV range variability is not solely a climate phenomenon but is substantially shaped by traffic conditions and road-network characteristics, factors that are amenable to policy intervention through congestion management and route optimisation. While the specific coefficients are derived for a single vehicle model under UK conditions, the framework and decomposition methodology are designed for broader application through re-parameterisation.

\section*{Acknowledgements}

The authors gratefully acknowledge King’s College London for providing the financial support for Liang Zhang’s PhD project.

\bibliographystyle{elsarticle-num}
\bibliography{references}

\end{document}